\documentclass[aps,showpacs,onecolumn,superscriptaddress]{revtex4}
\usepackage{amsmath,graphics}
\usepackage[next]{inputenc}
\usepackage{graphicx}
\usepackage{hyperref}
\usepackage{color}
\def\be{\begin{equation}}
\def\ee{\end{equation}}

\def\bi{\begin{itemize}}
\def\ei{\end{itemize}}
\def\bn{\begin{enumerate}}
\def\en{\end{enumerate}}
\def\bea{\begin{eqnarray}}
\def\eea{\end{eqnarray}}
\def\no{\nonumber}
\def\ba{\begin{array}}
\def\ea{\end{array}}
\def\bd{\begin{displaymath}}
\def\ed{\end{displaymath}}

\begin{document}

\title{Topological Phase Transition in the Extended Cluster Compass Ladder}
\author{R. Jafari}
\affiliation{ Research Department, Nanosolar System Company (NSS), Zanjan 45158-65911, Iran}
\email[]{jafari@iasbs.ac.ir, rohollah.jafari@gmail.com}
\affiliation{ Department of Physics, Institute for Advanced
Studies in Basic Sciences (IASBS), Zanjan 45137-66731, Iran}
\author{S. Mahdavifar}
\affiliation{Department of Physics, University of Guilan, P.O.Box 41335-1914, Rasht, Iran}
\email[]{mahdavifar@guilan.ac.ir}

\begin{abstract}
We have studied the exact solution of the extended cluster compass ladder,
which is equivalent to extended quantum compass model with cluster interaction
between next-nearest-neighbor spins, by using the Jordan-Wigner transformation.
We show that this model is always gapfull except at the critical surfaces where
the energy gap disappears. We obtain the analytic expressions of all critical couplings
which drive quantum phase transitions.
This model shows a rich phase diagram which includes spin-flop, strip antiferromagnetic
and topological ordered on the legs, in addition to the phase
with antiparallel ordering of spin $y$ component on the rungs. We study also the universality
and scaling properties of the three point correlation functions derivatives
in different regions to confirm the results obtained using the energy gap analysis.
On the other hand, we have replaced the cluster interaction with usual form and using
the Lanczos method a numerical experiment is done. Analyzing the numerical results,
we show that the effect of the cluster interaction between next-nearest-neighbor spins is completely
different from the usual form.
\end{abstract}
\date{\today}

\pacs{75.10.Pq, 64.70.Tg, 03.67.Mn, 75.10.Jm}

\maketitle


\section{Introduction \label{introduction}}

Ladder systems are well known for their many novel
properties and their relative simplicity makes them an ideal candidate
for much theoretical works \cite{Dagotto,Maekawa}.
Several experimental systems are known to be of or dominated by a
ladder-type structure, and theoretical studies have been
able to make reasonable predictions about the phases,
symmetries and transport properties of these materials \cite{Scalapino,Mayaffre,Blumberg}.
In particular, quantum spin ladders with frustration for half-integer and integer
spins, set up an important part of the researches since they present a unique testing
ground based on the available powerful analytical and numerical approaches for
one-dimensional (1D) systems. Specially the frustrated ladder models have
allowed controlled calculations to examine the topological order \cite{White}, the dimer order \cite{Starykh,Jafari3,Jafari4, Mahdavifar08, Vahedi12},
as well as the appearance of fractional excitations in spin models \cite{Allen}.
Although the exact results for frustrated systems are still
limited (see Ref. [\onlinecite{Miyahara}] and references cited therein),
the subject of integrable or exactly solved models in statistical mechanics
is quite important in both physics and mathematics since they provide a
rigorous information about the complex behavior of frustrated models.

\begin{figure}[t]
\begin{center}
\includegraphics[width=8.5cm]{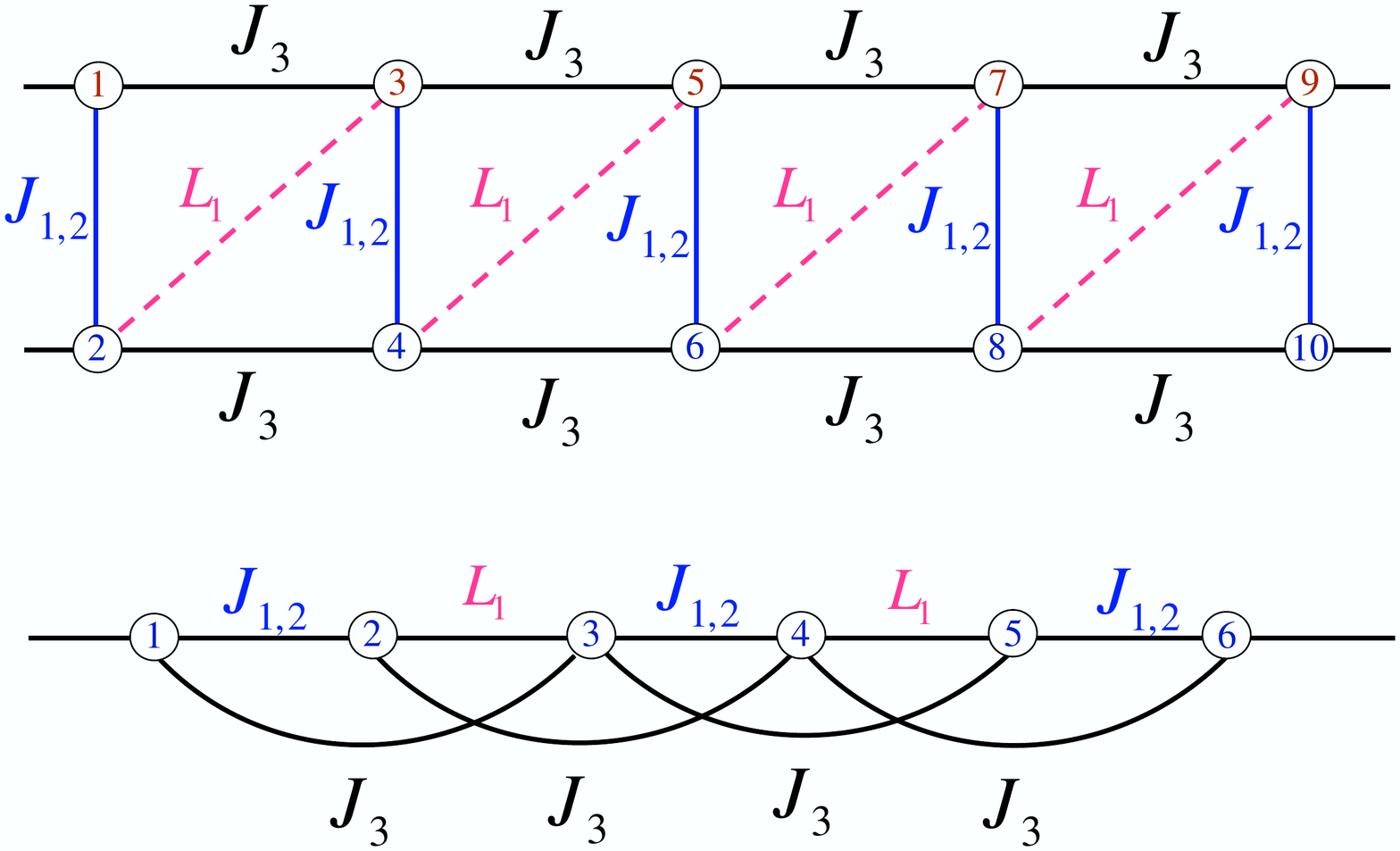}
\caption{(Color online) Schematic view of the extended quantum compass
ladder which is equivalent to a extended quantum compass model
with the cluster interaction between NNN spins. Interactions along
the ladder legs labeled as $J_{3}$. The interactions
along the rungs labeled as $J_{1,2}$ and $L_{1}$.} \label{fig1}
\end{center}
\end{figure}

In the other hand, building on the deep understanding of the Heisenberg and other models of magnetism,
it is a very common practice to describe discrete
degrees of freedom as pseudospins, with the hope to gain insight from the form of
the resulting magnetic model. A well-known example of considerable current interest shows up
in the context of Mott insulators with orbital degeneracy. A simplified model which describes
the nature of the orbital states in the case of a twofold degeneracy is the Quantum
Compass (QC) Model \cite{Kugel}. For simplicity, the 1D QC model, is
constructed by antiferromagnetic order of $X$ and $Y$ pseudospin components on odd and even bonds,
respectively \cite{Brzezicki,Brzezicki2,Nussinov,Eriksson,Mahdavifar,Jafari1,Motamedifar,Jafari2, Motamedifar13}.
The 1D QC model in the presence of a transverse field is exactly solvable
using the Jordan-Wigner transformation and exhibits interesting properties while
approaching to the quantum critical point at zero temperature \cite{Jafari1, Jafari2}.
In addition, its ladder version is solvable and its partition function can be obtained
exactly in case of a large (but finite) system \cite{Feigelman}.
The exact solution of compass two-leg ladder is exactly solvable by mapping to quantum Ising model
and exhibits interesting properties \cite{Brzezicki3}. However studying the compass model on the square
lattice using the exact diagonalizations, Greens function Monte Carlo simulations
and high-order perturbation theory prove that the model exhibits finite-temperature Ising transition
between $x$ and $z$ part of the Hamiltonian \cite{Dorier}.
To the best of our knowledge, the extended compass zigzag ladder and extended cluster
compass ladder (ECCL) which is equivalent to a QC model with
three-spin interaction between next-nearest neighbor (NNN) spins (see Fig.(\ref{fig1})) has not been
studied so far.
It is quite intresting to mention that Hamiltonian with three-spin
interaction so-called cluster interaction can be procreated using
optical lattices \cite{Pachoes}, which has been shown to
play an important role as a resource in the context of quantum
computation \cite{Raussendorf, Son, Smacchia, Orus, Kalis}.

In this paper, we obtain the exact solution of ECCL by using
the Jordan-Wigner (JW) transformation.
We show that this model reveals a rich phase diagram which includes quantum critical surfaces
depending on exchange couplings.
Moreover, because of nice scaling properties of correlation functions,
we will study  the divergence and scaling properties of three point correlation
(TPC) function near the quantum critical points (QCP). However, we have considered the
extended compass zigzag ladder (QC model with added usual NNN interaction) which is not exactly solvable, and we have
studied the magnetic induced effects of the usual NNN interaction on the ground state
phase diagram of the QC model using the numerical Lanczos method.
Based on the numerical results, we show that the effect of the cluster interaction between NNN
spins is completely different from the usual form.

The paper is organized as follows: In the next section, the
model is introduced and the exact solution is
determined analytically. In the section III,
the ground state phase diagram is obtained.
In the section IV, the universality and  scaling behavior of
the TPC function is investigated. In section V,
the results of a numerical simulation on the ground state
phase diagram of the QC model with added usual NNN interaction are presented. Finally, we
will discuss and summarize our
results in section VI.

\section{Hamiltonian and Exact Solution\label{EQCMTF}}

The Hamiltonian of a 1D cluster compass ladder is given by
\bea
\no
H=\sum_{n=1}^{N'}&[&J_{1}\sigma^{x}_{2n-1}\sigma^{x}_{2n}+
J_{2}\sigma^{y}_{2n-1}\sigma^{y}_{2n}+ L_{1}\sigma^{x}_{2n}\sigma^{x}_{2n+1}
\label{eq1}
+J_{3}(\sigma^{x}_{2n-1}\sigma^{z}_{2n}\sigma^{x}_{2n+1}+\sigma^{x}_{2n}\sigma^{z}_{2n+1}\sigma^{x}_{2n+2})],
\label{Hamiltonian}
\eea
where $J_{1}$ and $J_{2}$ are the odd bond exchange couplings, $L_{1}$ is the even bond exchange coupling while
$J_{3}$ denotes the strength of the cluster interaction and $N=2N'$ is the number of spins.  We assume periodic boundary conditions.
Note that the Hamiltonian is invariant under $J_{n}, L_{1}\longrightarrow -J_{n}, -L_{1},(n=1,2)$.
We can understand this by noting that a $\pi$-rotation around z axis on odd (or even) sites
maps $H(J_{1},J_{2},L_{1},J_{3})$ to $H(-J_{1},-J_{2},-L_{1},J_{3})$.
This is a consequence of the $Z_{2}\times Z_{2}$ symmetry of the cluster
state implemented precisely by $U_{1}$ and $U_{2}$ \cite{Smacchia}.
Therefore, without loss of generality we can restrict ourselves to $J_{2},L_{1},J_{3}\geq 0$.
In order to diagonalize the Hamiltonian we shall first express the Hamiltonian
(Eq. (\ref{eq1})) in terms of fermion operators.
This can be done in usual way applying the Jordan-Wigner transformation \cite{Brzezicki2,Jordan,Perk}
as defined below,

\bea
\no
\sigma^{x}_{j}=b^{+}_{j}+b^{-}_{j},~~
\sigma^{y}_{j}=b^{+}_{j}-b^{-}_{j},~~
\sigma^{z}_{j}=2b^{+}_{j}b^{-}_{j}-1,~~
b^{+}_{j}=c^{\dag}_{j}~e^{i\pi\Sigma_{m=1}^{j-1}c^{\dag}_{m}c_{m}},~~
b^{-}_{j}=e^{-i\pi\Sigma_{m=1}^{j-1}c^{\dag}_{m}c_{m}}~c_{j},
\eea
which transforms spins into the fermion operators $c_{j}$.

The crucial step is to define the independent Majorana fermions
\cite{Brzezicki2,Perk,Sengupta} at site $n$, $c_{n}^{q}\equiv c_{2n-1}$
and $c_{n}^{p}\equiv c_{2n}$. This can be regarded as quasiparticles'
spin or as splitting the chain into bi-atomic elementary cells \cite{Brzezicki2,Perk}.
Substituting for $\sigma^{x}_{j}$, $\sigma^{y}_{j}$ and $\sigma^{z}_{j}$ ($j=2n, 2n-1$)
in terms of Majorana fermions with antiperiodic boundary condition (subspace with an
even number of fermions) followed by a Fourier transformation, Hamiltonian
Eq. (\ref{eq1}) (apart from additive constant), can be written in the diagonal
block form

\bea
\label{eq2}
H=\sum_{k}\Gamma^{\dag}_{k}.A(k).\Gamma_{k},~~\Gamma^{\dag}_{k}=(c_{k}^{q},c_{-k}^{p\dag},c_{-k}^{q},c_{k}^{p\dag})
\eea

where
\begin{widetext}
\bea
\no
A(k)=\left(
       \begin{array}{cccc}
         -2J_{3}\cos(k) & J_{1}-J_{2}-L_{1}e^{ik} & 2iJ_{3}\sin k & J_{1}+J_{2}+L_{1}e^{ik} \\
         J_{1}-J_{2}-L_{1}e^{-ik} & 2J_{3}\cos(k) & -J_{1}-J_{2}-L_{1}e^{-ik} & -2iJ_{3}\sin k \\
         -2iJ_{3}\sin k & -J_{1}-J_{2}-L_{1}e^{ik} & 2J_{3}\cos(k) & -J_{1}+J_{2}+L_{1}e^{ik} \\
         J_{1}+J_{2}+L_{1}e^{-ik} & 2iJ_{3}\sin k & -J_{1}+J_{2}+L_{1}e^{-ik} & -2J_{3}\cos(k) \\
       \end{array}
     \right)
\eea
\end{widetext}
The matrix $A(k)$ can be diagonalized easily and we find the Hamiltonian Eq. (\ref{eq2}) in a diagonal form

\bea
\label{eq3}
H=\sum_{k}\Big[E^{q}_{k}(\gamma_{k}^{q\dag}\gamma_{k}^{q}-\frac{1}{2})+
E^{p}_{k}(\gamma_{k}^{p\dag}\gamma_{k}^{p}-\frac{1}{2})\Big],
\eea
where $E^{q}_{k}=\sqrt{2(a+c)}$, $E^{p}_{k}=\sqrt{2(a-c)}$ and $c=\sqrt{a^{2}-b}$ in which
\bea
\no
a=J_{1}^{2}+J_{2}^{2}+L_{1}^{2}+2J_{3}^{2}+2J_{2}L_{1}\cos k,~~
b=4[J_{3}^{4}+J_{1}^{2}(J_{2}^{2}+L_{1}^{2})]-8J_{1}L_{1}(J_{3}^{2}-J_{1}J_{2})\cos k
-8J_{1}J_{2}J_{3}^{2}\cos 2k.
\eea
The ground state ($E_{G}$) and the lowest excited state ($E_{E}$) energies are obtained from Eq.(\ref{eq3}),
\bea
\no
E_{G}=-\frac{1}{2}\sum_{k}(E^{q}_{k}+E^{p}_{k}),~~E_{E}=-\frac{1}{2}\sum_{k}(E^{q}_{k}-E^{p}_{k}),
\eea
which can be written as a function of $a$ and $b$,
\bea
\label{eq4}
E_{G}=-2\sum_{k>0}\sqrt{a+\sqrt{b}},~~E_{E}=-2\sum_{k>0}\sqrt{a-\sqrt{b}}.
\eea
The energy gap will occur at a wave vector $k_{0}$ that
\bea
\no
\frac{dE(k=k_{0})}{dk}=0.
\eea
The energy gap wave vector $k_{0}$ is given by

\bea
\label{eq5}
k_{0}=0,\pi,~\cos k_{0}=\frac{L_{1}(J_{1}J_{2}-J_{3}^{2})}{4J_{2}J_{3}^{2}}\pm \sqrt{\frac{L_{1}^{2}(J_{1}J_{2}+J_{3}^{2})(4J_{2}J_{3}^{2}+J_{1}L_{1}^{2})}
{16J_{1}J_{2}J_{3}^{2}(4J_{1}J_{3}^{2}+J_{2}L_{1}^{2})}},
\eea

\begin{figure}[t]
\begin{center}
\includegraphics[width=9cm]{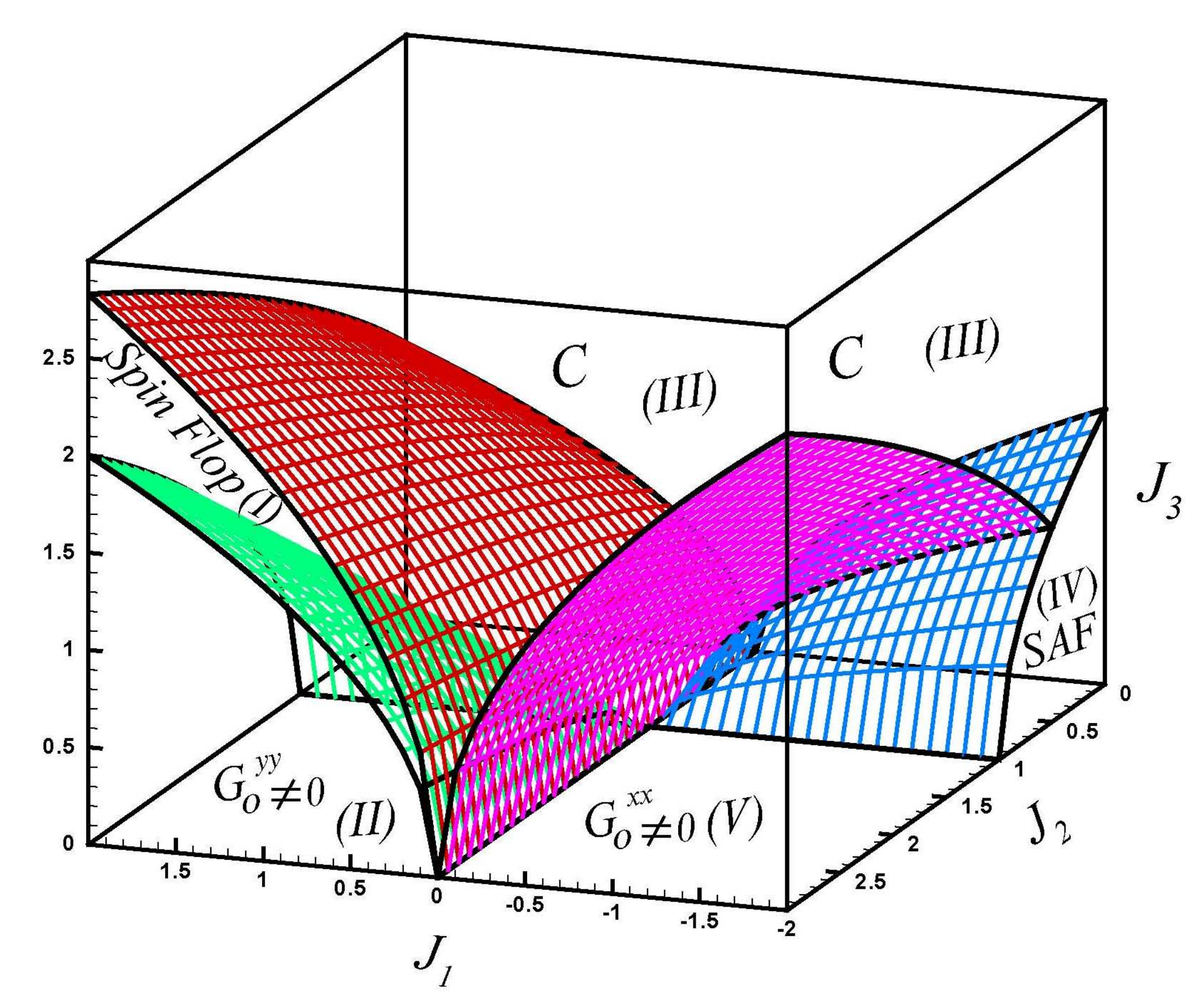}
\caption{(Color online) Phase diagram of the cluster compass ladder.
For $J_{1}>0, J_{2}<1$, the front and top sides of the green convex surface are spin-flop phase (I) and the back side ($J_{1}>0, J_{2}>1$) is specified by the antiparallel order of the spin $y$ component (II) on odd bonds.
The  red checkerboard pattern shows the boundary between the spin-flop phase (I) and cluster phase (III).
In the case of $J_{1}<0, J_{2}<\frac{1}{2}$ there are two phases, the strip antiferromagnetic phase (IV) which exists below the blue convex surface and the cluster phase (III) which is above it. For $J_{1}<0, J_{2}>\frac{1}{2}$ the system displays a cluster phase (III) above the blue and purple convex surfaces. The purple convex surface manifest boundary between parallel order of the spin $x$ component (V) on odd bonds and cluster phase (III).} \label{fig2}
\end{center}
\end{figure}

\begin{figure}
\begin{center}
\includegraphics[width=9cm]{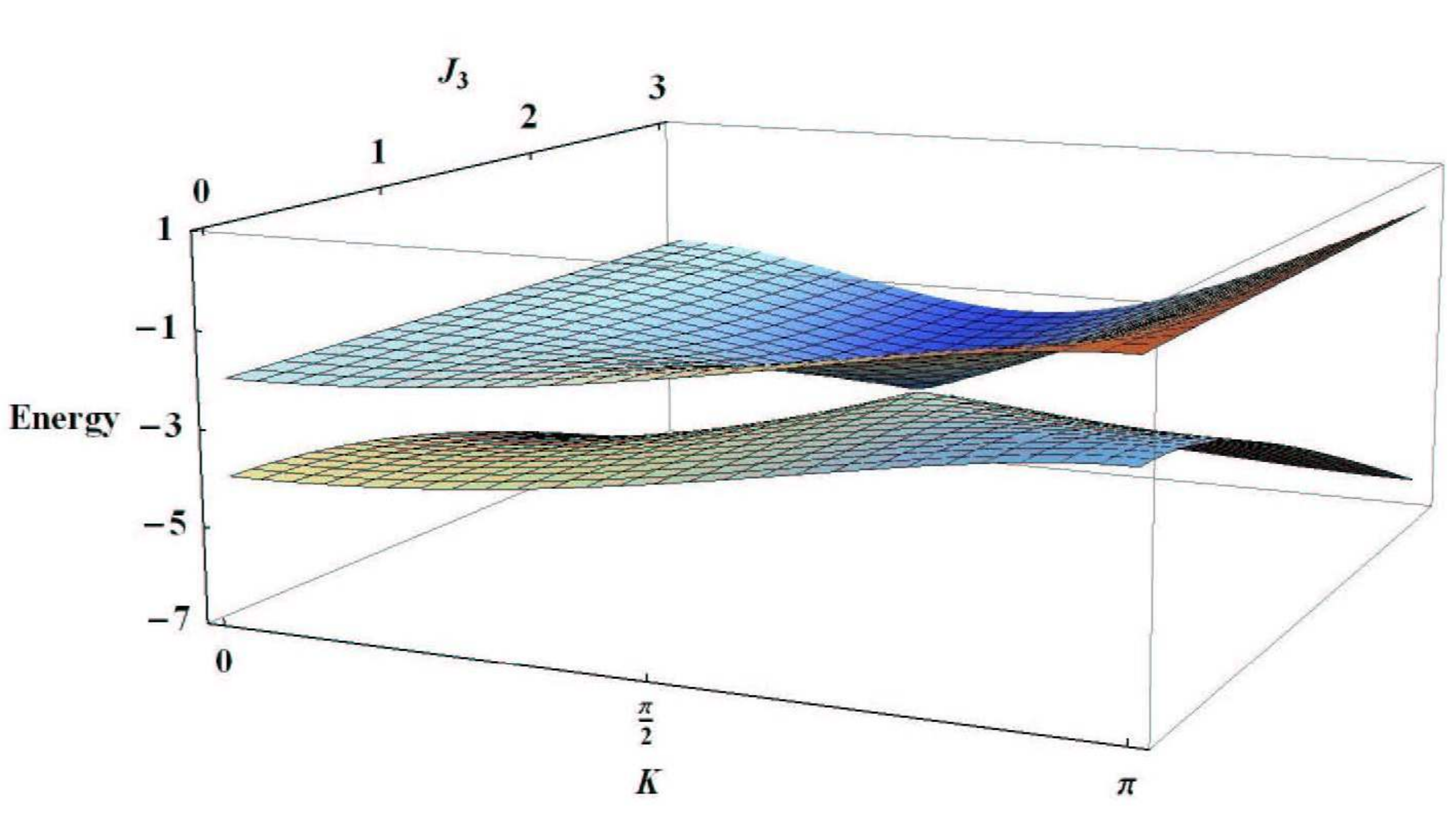}
\caption{(Color online) The ground and first excited states
versus $k$ and $J_3$ for $J_1=-1$ and $J_2=2$.} \label{fig3}
\end{center}
\end{figure}

The system is at criticality when the gap vanishes. It can be shown that the gap
of the spectrum vanishes at $J_{3}^{0}=\sqrt{J_{1}(J_{2}+L_{1})}$ and
$J_{3}^{\pi}=\sqrt{J_{1}(J_{2}-L_{1})}$ with ordering of wavevectors $k_{0}=0$ and $k_{0}=\pi$
respectively.
Moreover there is an additional phase transition at $J_{3}^{k_{0}}=-\imath\sqrt{J_{1}J_{2}}$
which shows the commensurate and incommensurate transition and it will show the effect
of disorder on the phases \cite{Nijs}. The phase boundary separates the commensurate phase
from the incommensurate phase (Cluster phase) with ordering wavevector given by Eq. (\ref{eq5}).
The incommensurate wavevector pick up a value $\cos k_{0}=-L_{1}/2J_{2}$ (Fig. (\ref{fig3})) at the phase boundary.
So, the quantum phase transition (QPT) which could be
driven by cluster interaction, depending on exchange couplings, occurs
at $J_{3}^{0}$, $J_{3}^{\pi}$ and $J_{3}^{k_{0}}$.

\begin{figure*}
\centerline{\includegraphics[width=9cm]{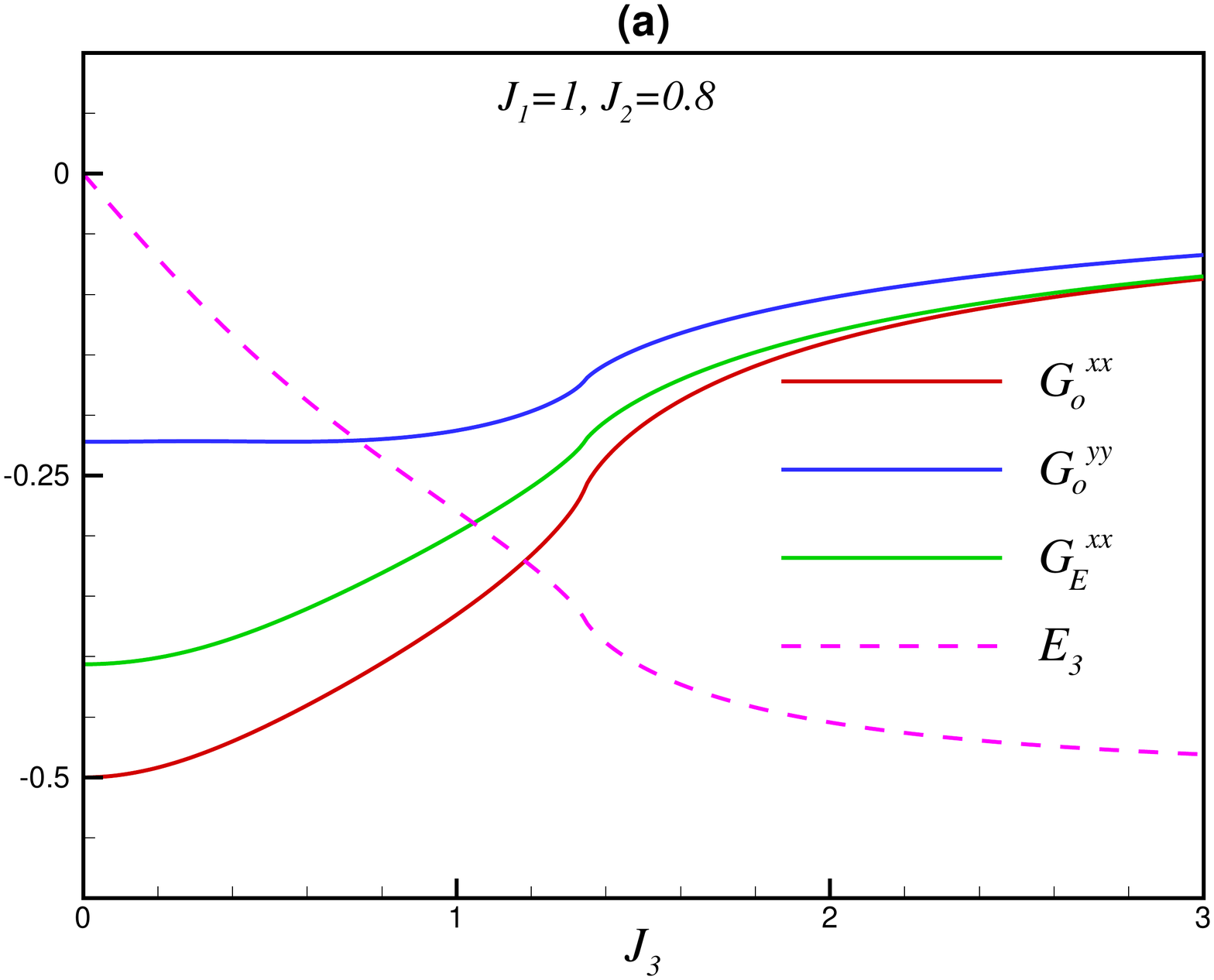}
\includegraphics[width=9cm]{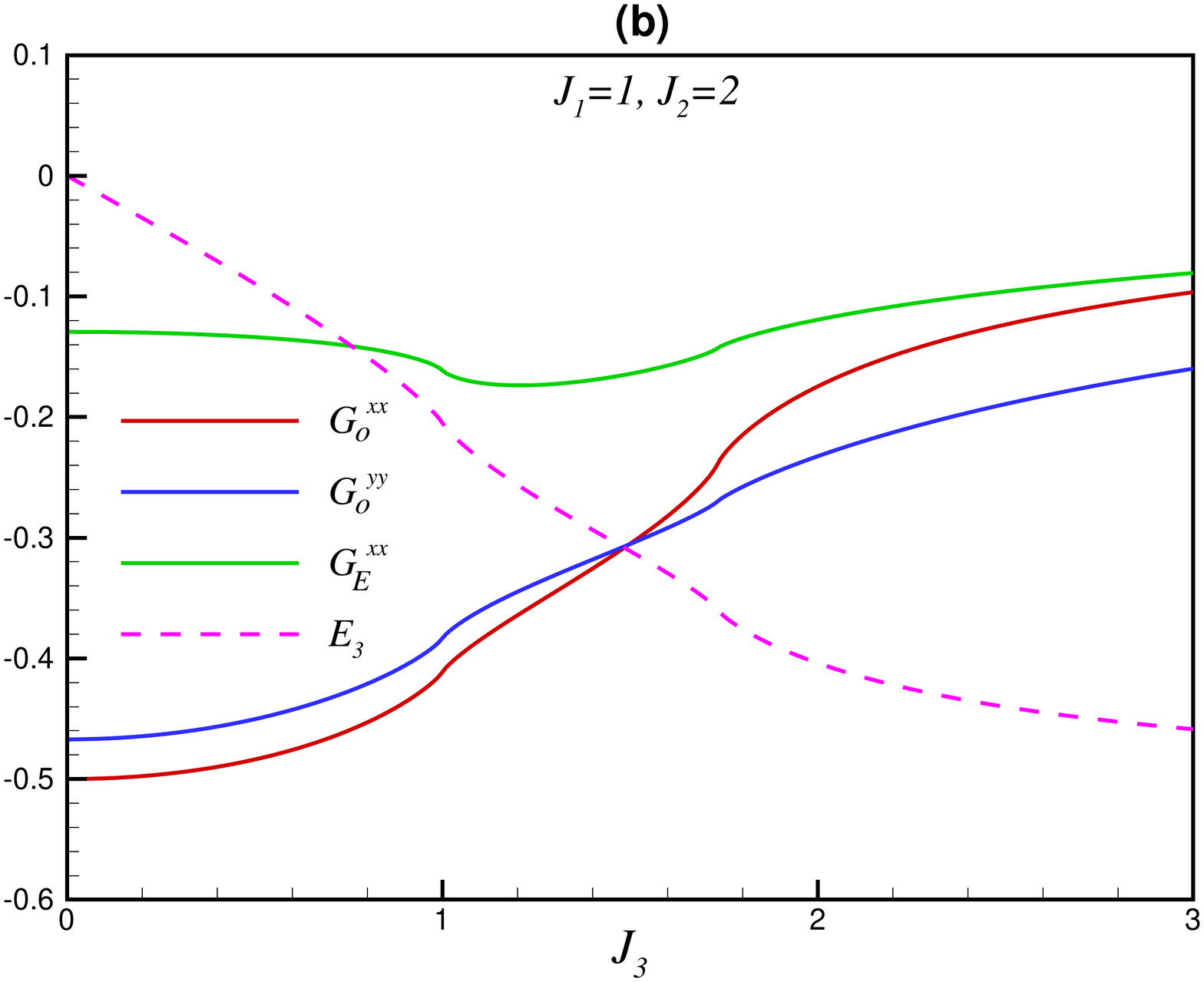}}
\caption{(Color online.) The components of the nearest-neighbor
spin correlation functions on even and odd bonds and
$E_{3}$ for (a) $J_{1}=1, J_{2}=0.8$ and (b) $J_{1}=1, J_{2}=2$.}\label{fig4}
\end{figure*}


\section{Phase Diagram\label{PD}}

For $J_{3}=0$ (the interactions along the legs are zero) the model has been decoupled to two
1D extended compass model. The complete phase diagram of the extended compass model has been reported in
Refs. [\onlinecite{Eriksson,Mahdavifar}] and [\onlinecite{Jafari1}]. They have shown that the first-order transition occurs at the multicritical point where a line of first-order transition ($J_{1}/L_{1}=0$) meets with a line of second order transition ($J_{2}/L_{1}=1$). Also, there are four gapped phases in the exchange couplings' space,

\begin{itemize}
  \item (I) $J_{1}>0,~J_{2}<1$: In this region the ground state is in the Ne\'{e}l phase along the $x$ axis.
  \item (II) $J_{1}>0,~J_{2}>1$: In this case there is antiparallel ordering of spin $y$ component on odd bonds.
  \item (III) $J_{1}<0,~J_{2}>1$: In this case there is parallel ordering of spin $x$ component on odd bonds.
  \item (IV) $J_{1}<0,~J_{2}<1$: In this region the ground state is in the strip antiferromagnetic (SAF) phase.
 \end{itemize}

For $J_{1}=J_{2}=L_{1}=0,$ the ground state is a cluster state \cite{Hein}.
The Phase diagram of the cluster compass ladder is shown in Fig. (\ref{fig2}).
Depending on the exchange couplings, the cluster exchange could result phase transitions at
$J_{3}^{0}$, $J_{3}^{\pi}$ and $J_{3}^{k_{0}}$ where the energy gap vanishes (For simplicity we take $L_{1}=1$).
The two point nearest-neighbor (NN) correlation functions ($G_{0}^{\alpha\alpha}, \alpha=x, y$) and TPC function ($E_3$) are incidentally the expectation values of the coupling terms in the Hamiltonian,

\begin{figure*}
\centerline{\includegraphics[width=9cm]{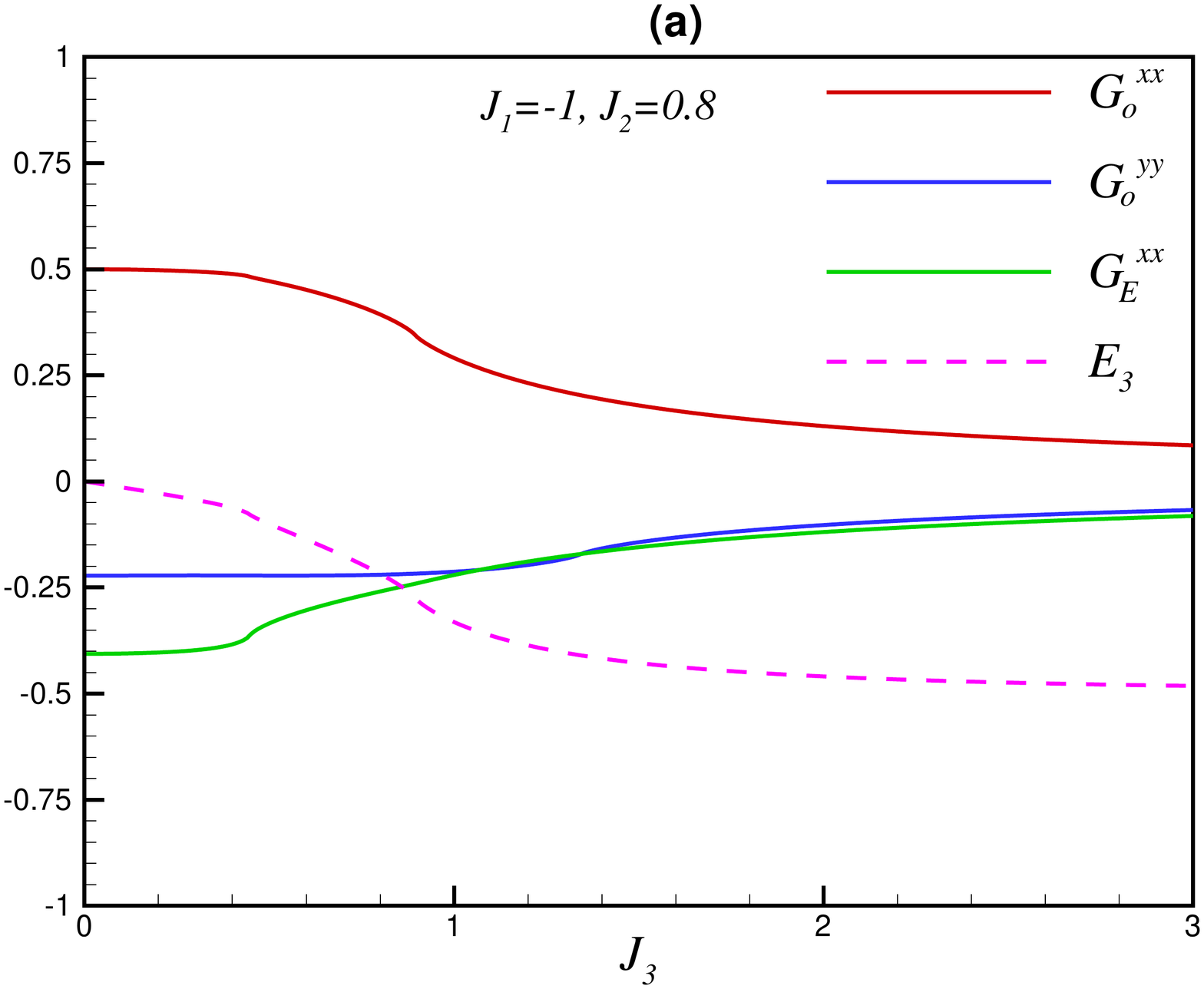}
\includegraphics[width=9cm]{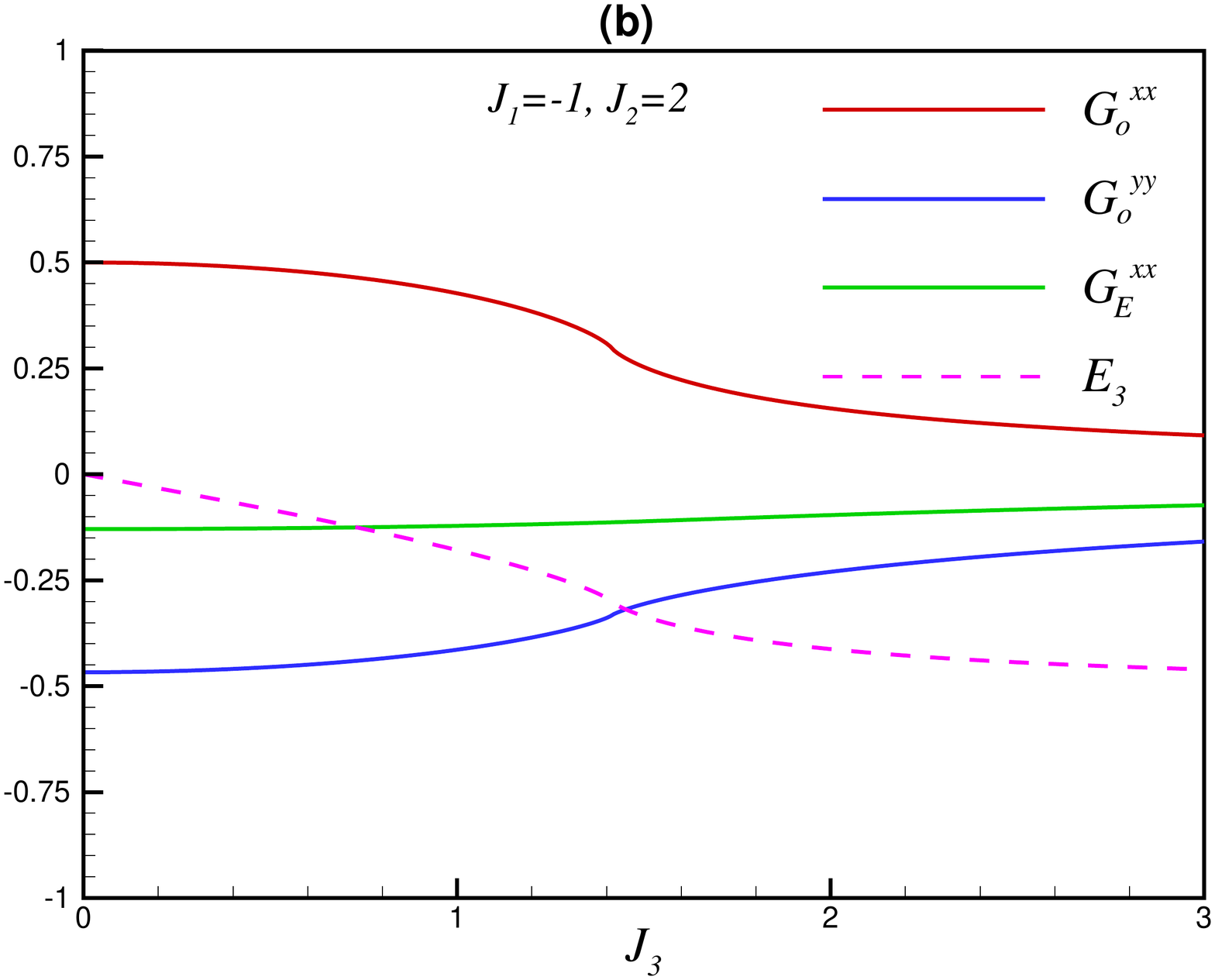}}
\caption{(Color online.) The different components of the nearest-neighbor
spin correlation functions on even and odd bonds and
$E_{3}$ for (a) $J_{1}=-1, J_{2}=0.8$ and (b) $J_{1}=-1, J_{2}=2$.}\label{fig5}
\end{figure*}

\bea
\no
G^{xx}_{o}=<\sigma^{x}_{2i-1}\sigma^{x}_{2i}>=\frac{dE_{G}}{dJ_{1}},
~G^{yy}_{o}=<\sigma^{y}_{2i-1}\sigma^{y}_{2i}>=\frac{dE_{G}}{dJ_{2}},
~G^{xx}_{E}=<\sigma^{x}_{2i}\sigma^{x}_{2i+1}>=\frac{dE_{G}}{dL_{1}},
~E_{3}=<\sigma^{x}_{1}\sigma^{z}_{2}\sigma^{x}_{3}>=\frac{dE_{G}}{dJ_{3}}
\eea

Fig. \ref{fig4}(a) shows NN correlation functions on odd and even bonds and $E_{3}$
for an infinite system size in the region (I) ($J_{1}=1, J_{2}=0.8$).
In this region tuning the cluster exchange forces the system
to fall into a topological (cluster) phase. The spin-flop-topological phase transition occurs
at $J_{3}^{c}=J_{3}^{0}=\sqrt{1.8}$ (red checkerboard curved plane in Fig. (\ref{fig2}))
under which surface the ground state is in the spin-flop phase (the Ne\'{e}l order along the x axis).
It is seen in Fig.~\ref{fig4}(a) that the onset of the cluster exchange sets up the $E_{3}$
immediately, and its absolute value continuously increases with an increase in $J_{3}$ to its saturated value.
However, the antiparallel order of the $x$ and $y$ spin components on odd ($G^{xx}_{o}, G^{yy}_{o}$)
and even ($G^{xx}_{E}$) bonds are reduced by increasing the $J_{3}$ and goes to zero for $J_{3}\rightarrow\infty$.

In  the region (II) the gap decreases with increasing the cluster exchange and
disappears at the lower critical point $J_{3}^{c_{1}}=J_{3}^{\pi}=1$
(green checkerboard curved plane in Fig. (\ref{fig2})). Beyond this critical point
the energy gap immediately appears with the increasing the cluster exchange and this
process continues until the upper critical field $J_{3}^{c_{2}}=J_{3}^{0}=\sqrt{3}$
(red checkerboard curved plane in Fig. (\ref{fig2})) is reached at which the energy gap vanishes.
The model becomes once again gapped above the second critical point $J_{3}^{c_{2}}=\sqrt{3}$.
Fig.~\ref{fig4}(b) shows the NN correlation functions and $E_{3}$  versus the cluster exchange
in the region (II) ($J_{1}=1, J_{2}=2$). It manifests that under the lower critical
point ($J_{3}<J_{3}^{c_{1}}=1$) the antiparallel order of the $x$ spin component on the
even bond stay quite unchanged while the spin $y$ and $x$ components on the odd bond decreases
continuously with an increase the cluster interaction. $E_{3}$ increase gradually
as $J_{3}$ increases and tend to saturate value for $J_{3}\rightarrow\infty$.
Above the $J_{3}^{c_{1}}=1$ the antiparallel ordering of the $x$ and $y$ spin components on
the odd bound decreases gradually as $J_{3}$ increases and tend to zero above the
$J_{3}^{c_{2}}=J_{3}^{0}=\sqrt{3}$.

The result is interesting in the intermediate region of the cluster exchange
$J^{c_{1}}_{3}<J_{3}<J^{c_{2}}_{3}$ where increasing the $J_{3}$ enhances the
antiparallel ordering of $x$ spin component on even bonds up to a maximum and
then decreases gradually, while enhancing the cluster exchange decreases the
other antiparallel ordering. So we predict that the gapped spin-flop phase exists
in the intermediate values of the cluster exchange $J^{c_{1}}_{3}<J_{3}<J^{c_{2}}_{3}$.
In other words, in the region (II), the cluster exchange destroys the ground state's
antiparallel ordering of $y$ spin component on even bonds at $J^{c_{1}}_{3}$ and
forces the system into the spin-flop phase above the $J^{c_{1}}_{3}$.
The spin-flop-topological transition occurs beyond $J^{c_{2}}_{3}$.

Fig. \ref{fig5}(a) shows the $E_{3}$ and NN Correlation functions in the region
(IV) ($J_{1}=-1, J_{2}=0.8$).
This region includes two gapped phases, SAF and topological where are separated
from each other at the critical points $J^{c}_{3}=J^{\pi}_{3}=\sqrt{0.2}$
(blue checkerboard curved plane in Fig. (\ref{fig2})).

\begin{figure*}
\centerline{\includegraphics[width=9cm]{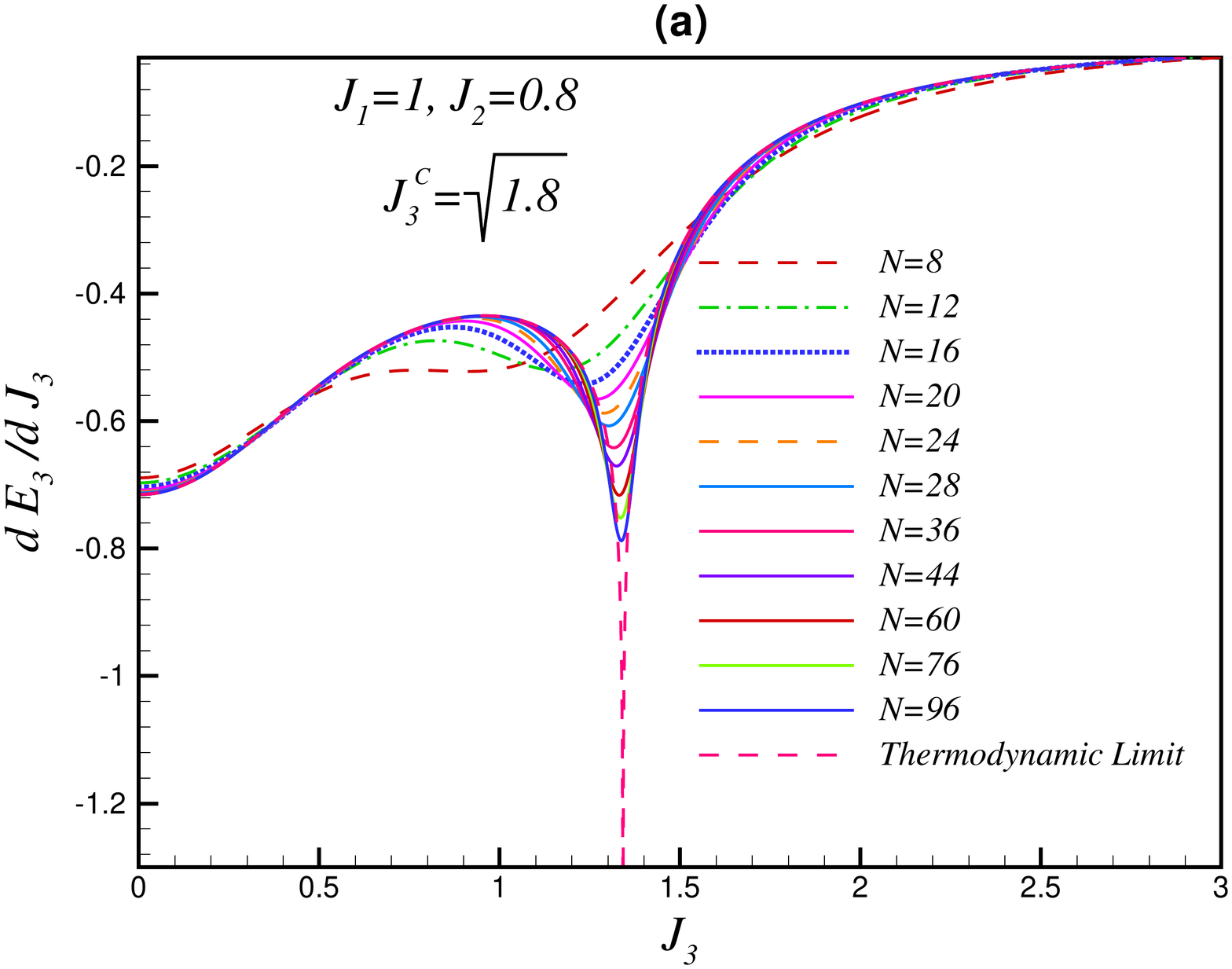}
\includegraphics[width=9cm]{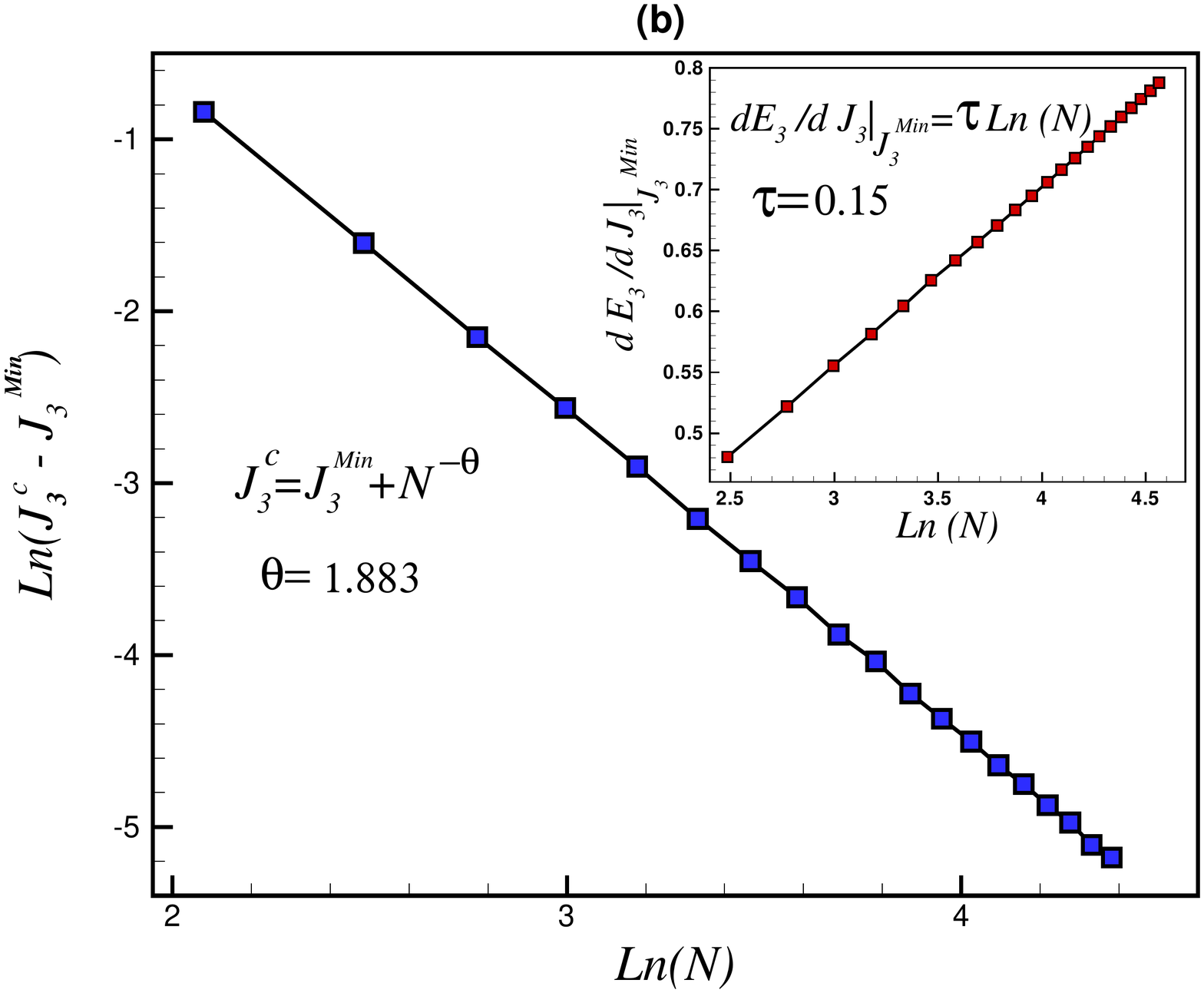}}
\caption{(Color online.) (a) The derivative
of $E_{3}$ in the region $J_1>0, J_2<1$ versus $J_3$ for different
chain lengths. (b) The scaling behavior of the $J_{3}^{c}-J_{3}^{Min}$
in respect to the chain length. (b) Inset: The finite-size scaling analysis for
the case of logarithmic divergence around the minimum value of the
derivative of $E_{3}$.} \label{fig6}
\end{figure*}

\begin{figure*}
\centerline{\includegraphics[width=9cm]{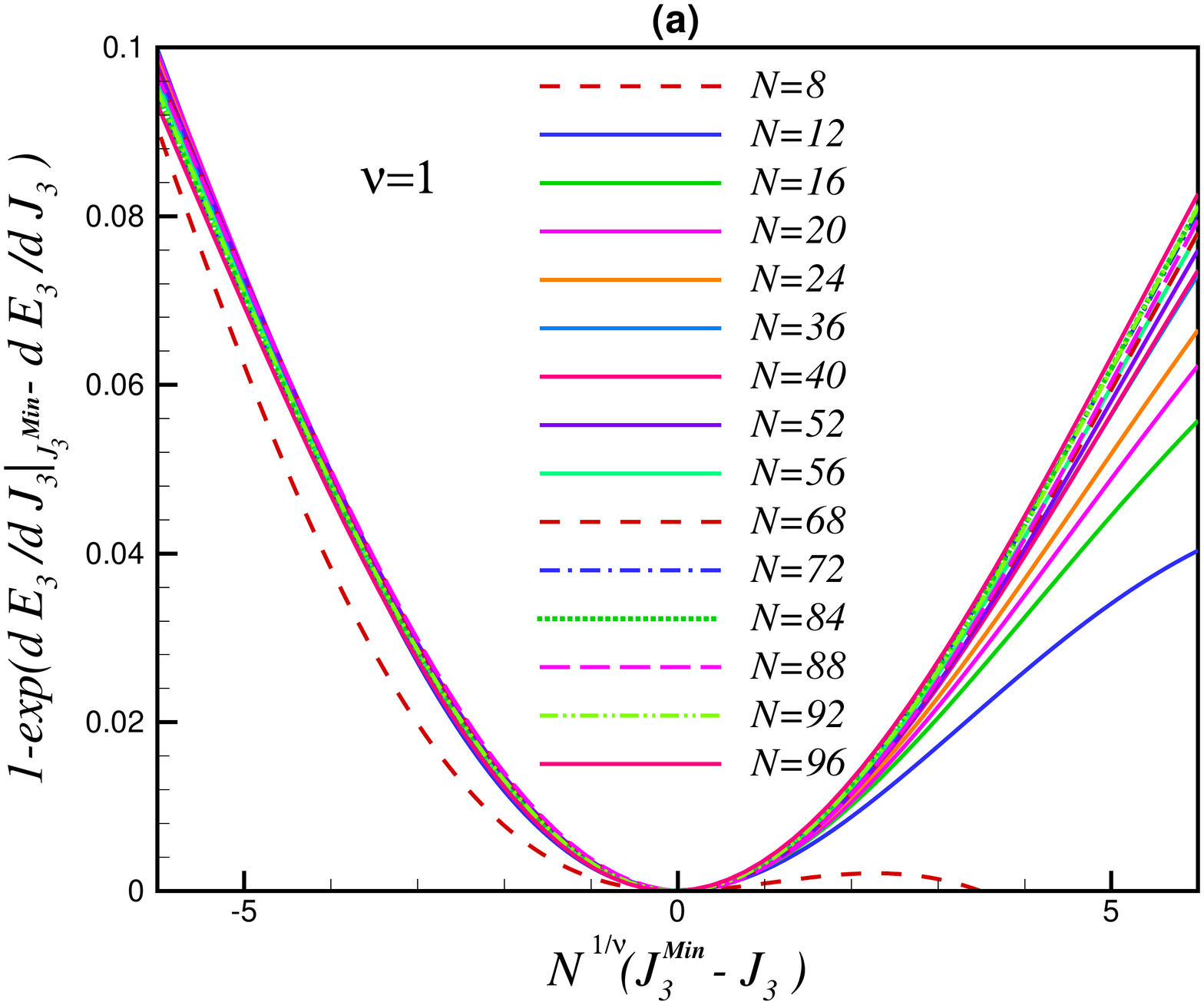}
\includegraphics[width=9cm]{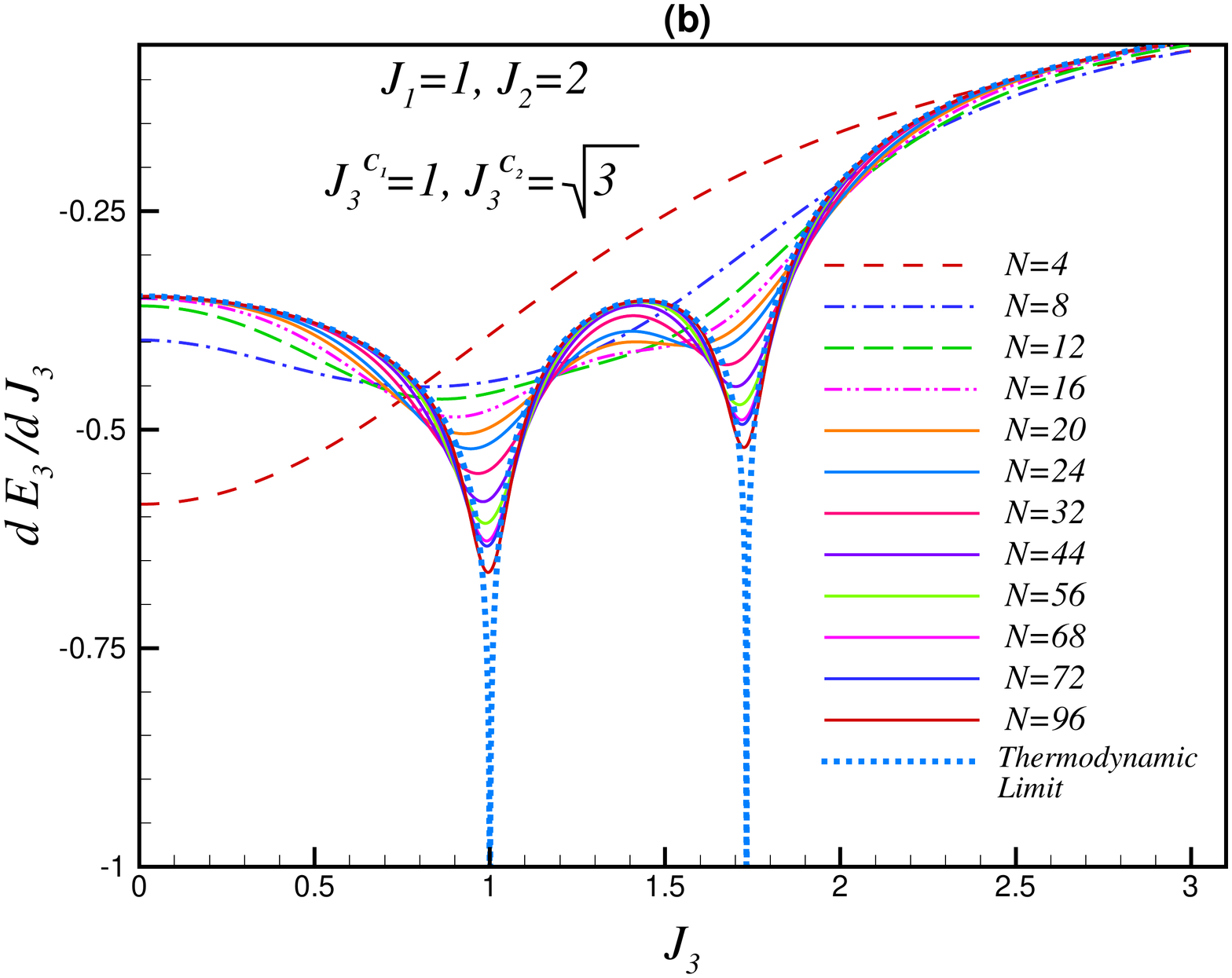}}
\caption{(Color online.) (a) The finite-size scaling of $dE_{3}/dJ_{3}$
for different lattice sizes. The curves which correspond
to different system sizes clearly collapse on a single curve.
(b) The first-order derivative of $dE_{3}/dJ_{3}$ as a
function of $J_{3}$ for various system sizes for
$J_{1}=-1, J_{2}=2$.} \label{fig7}
\end{figure*}

One of the interesting features of this model is the existence
of phases that appear because of the competition between the
usual term and cluster terms. A surprising result occurs
in $J_{1}<0$ and $J_{3}^{k_{0}}=\sqrt{|J_{1}|J_{2}}$ where gap
vanishes and there is a topological phase transition.
The $E_{3}$ and NN correlation functions have been depicted in Fig. \ref{fig5}(b) for
$J_{1}=-1, J_{2}=2$. As it is clear the spin components decreases with an
increase the cluster interaction and $E_3$ increase as $J_3$ increases but
there is a kink on the spin $x$ and $y$ components on the odd bond and
cluster interaction. The purple convex surface shows the topological phase transition
surface under which surface the ground state is in the parallel order of the spin $x$
component on odd bonds (V) and topological phase (III) which is above it.
In this region the derivatives of spin components (except $G_{E}^{xx}$) and cluster
interaction with respect to $J_3$ show the divergence in this region.

\section{Universality and scaling of Correlation Functions\label{USCF}}

The finite size scaling method is an efficient way for extracting critical exponents
from finite-size systems results. In this method one should compare a sequence of
finite lattices. The finite lattice systems are solved exactly, and various quantities
can be calculated as a function of the lattice size $N$, for small values of $N$.
Finally, these functions are scaled up to $N\longrightarrow\infty$.
In relation to this topic, an interesting topic is the study of non-analytic
behavior and finite size scaling of two-point correlation function\cite{Jafari1}.
In principle, two-point correlation functions show the universality and scaling
around the QCP and could capture QCP and also could reveal the scaling and
universality of entanglement near the QCP. In this section we will study the
behavior of TPC function derivative with respect to the cluster interaction.
Figure \ref{fig6}(a) shows the result of the TPC function derivative in the
region ($J_1>0$, $J_2<1$). It can be seen that the $dE_3/dJ_3$ for different $N$
all show a drop with a bit different position at the pseudo-critical point $J_3^{Min}$ in
where $dE_3/dJ_3$ becomes more pronounced by increasing $N$.
It is also deduced that, at the QCP the TPC function derivative, $dE_3/dJ_3$,
is an extensive quantity and diverges for the infinite chain.
In the other hand, by increasing $N$, the pseudo-critical point comes close to
the real critical point $J_3^{c}$. It is suggested a scaling behavior as

\bea
\no
|J_3^{Min}-J_3^{c}| \sim N^{-\theta}.
\eea

In Fig. \ref{fig6}(b), the value of $\ln(J_3^{c}-J_3^{Min})$ is plotted as a
function of $\ln(N)$. The best linear fit to our data is obtained with $\theta=1.883\pm 0.01$.
Moreover, we have derived the scaling behavior of $|dE_3/dJ_3|_{J_3^{Min}}$ versus $N$.
It is suggested a scaling behavior as $|dE_3/dJ_3|_{J_3^{Min}} \sim \ln N^{\tau}$.
The results are plotted in the inset of Fig.\ref{fig6}(b), which shows the linear behavior of
$|dE_3/dJ_3|_{J_3^{Min}}$ versus $\ln(N)$. The best fit is obtained with
$\tau=0.15\pm 0.01$.
According to the divergence behavior of the TPC function derivative at the critical
point $J_3^{c}$, the $|dE_3/dJ_3|$ in the thermodynamic limit
$N\longrightarrow \infty$ and in vicinity of $J_3^{c}$ behaves as

\begin{eqnarray}
|dE_3/dJ_3| \sim |J_3-J_3^{c}|^{-\nu}.
\end{eqnarray}

However, correspond to the scaling ansatz\cite{Barber83}, it is more convenient
to write the TPC function derivative in a finite size system as

\begin{eqnarray}
dE_3/dJ_3-dE_3/dJ_3|_{J_3^{Min}} \sim F(N^{1/\nu}(J_3-J_3^{Min}),
\end{eqnarray}
where $F(x)$ is known as the scaling function. To find the critical exponent $\nu$,
we have plotted $1-\exp(dE_3/dJ_3-dE_3/dJ_3|_{J_3^{Min}})$ versus the scaling variable
$N^{1/\nu}(J_3-J_3^{Min})$ in Fig. \ref{fig7}(a). The curves which correspond to different
chain sizes clearly collapse on a single universal curve with $\nu=1.00\pm 0.01$.
Which is exactly the same as the critical exponent of the correlation length of ITF model.

We have also investigated the behavior of the TPC function derivative in other regions.
The results in the region ($J_1>0$, $J_2>1$) are very interesting. As mentioned with increasing
the cluster interaction, two quantum phase transitions will be happened at $J_3^{c_1}$ and $J_3^{c_2}$.
The signature of these critical points clearly seen in Fig. \ref{fig7} (b).
It can be seen that the $dE_3/dJ_3$ for different $N$ all show two drops with a bit different
position at the pseudo-critical points $J_3^{Min_{1}}$ and $J_3^{Min_{2}}$.
We did the same analysis and results are presented in Table I.
The results show that the quantum phase transitions in this region also take
in the universality class of the ITF model.

\begin{figure}
\includegraphics[width=9cm]{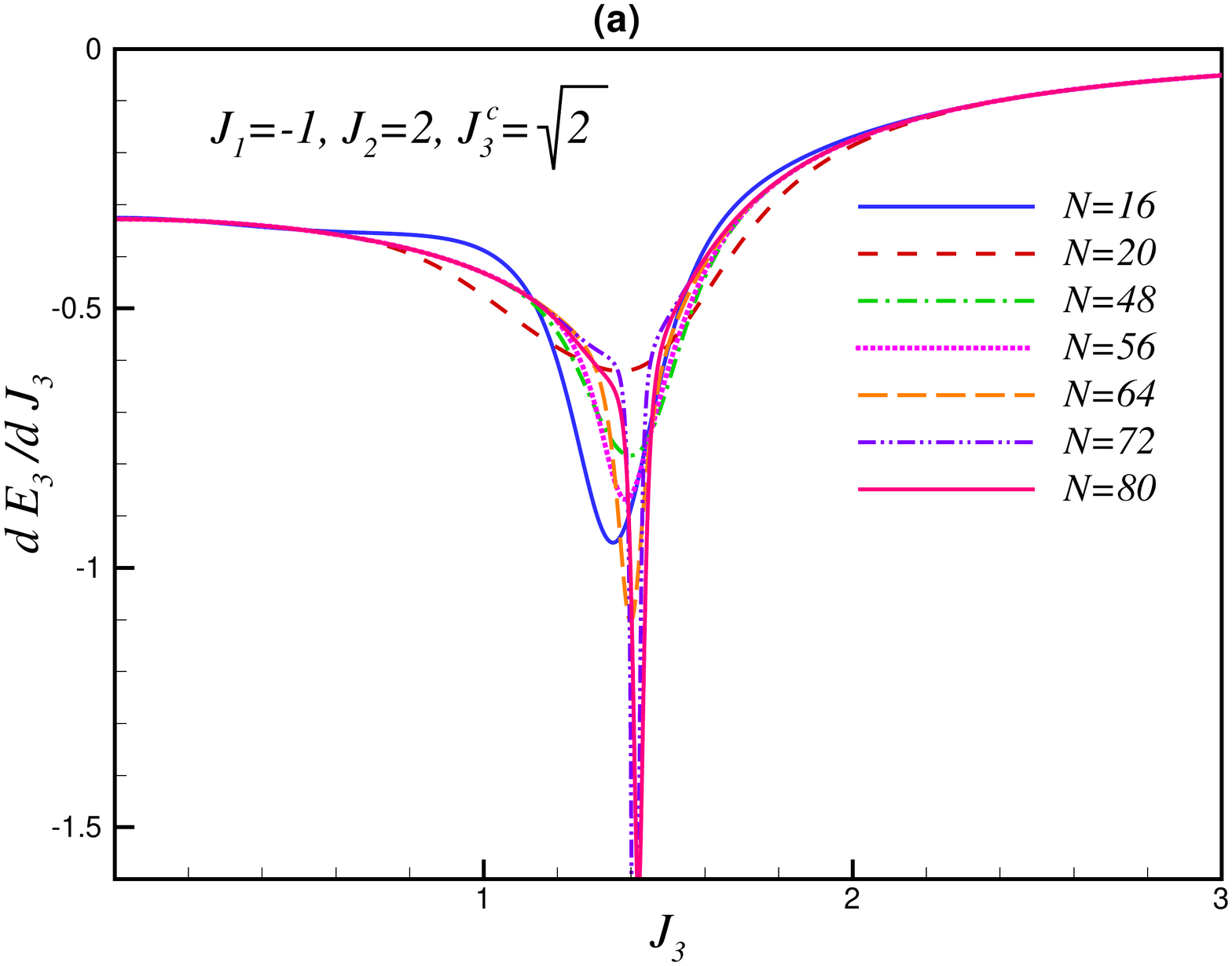}
\caption{(Color online.) Evolution of $dE_{3}/dJ_{3}$ versus $J_{3}$ for different
system sizes for $J_{1}=-1, J2=2$. (b) The numerical simulation results obtained from
the Lanczos for the string order parameter for finite chain length $N=12, 16, 20, 24$
for $J_{1}=-1, J_{2}=2$.} \label{fig8}
\end{figure}

As it is mentioned, in the region $J_{1}<0$ the derivatives of NN correlation functons
and TPC function with respect to $J_3$ show the divergence at the critical
point $J_{3}^{\pi},~J_{3}^{k_{0}}$, (Fig. (\ref{fig8})).
Unfortunately, unlike the scaling behavior of TPC function at the usual critical
surfaces ($J_{3}^{0},~J_{3}^{\pi}$), it does not show any scaling at the topological
phase transition surface ($J_{3}^{k_{0}}$).

In Refs. [\onlinecite{Son,Smacchia}] it was shown that the cluster-Ising model
with open boundary conditions has a fourfold degenerate
ground state that possesses symmetry-protected topological
order \cite{Kou,Liu,Titvinidze,Pollmann}, reflecting
the existence of the edge states \cite{Cui}. Without symmetry, the cluster
phase is a (non topological) quantum spin liquid, since there
is a gap and no symmetry is spontaneously broken.
We expect that this model has similar
features, though the presence of the nontrivial phases transition surface
$J_{3}^{k_{0}}$, at which the ground state is double degenerate, makes the situation more complicated.
Although local order parameter does not exist to characterize
the topological phase, a promising route to the characterization of
topological phase transition is the study of the geometric entanglement\cite{Wei03,Wei05} in ECCL,

\be
\label{SOP}
\varepsilon=-\log_{2}[max~|\langle \phi(\theta)|\psi\rangle|^{2}],
\ee
where $|\psi\rangle$ is the ground state of the system and $|\phi(\theta)\rangle$ is the closest separable state and is defined as
\be
\label{SOP1}
| \phi(\theta)\rangle=\Pi_{j=1}^{N}(\cos(\theta_{j})+e^{i\varphi_{j}} \sin(\theta_{j})\sigma_{j}^{x})|\uparrow\rangle^{\bigotimes N}.
\ee

\begin{figure*}
\centerline{\includegraphics[width=9cm]{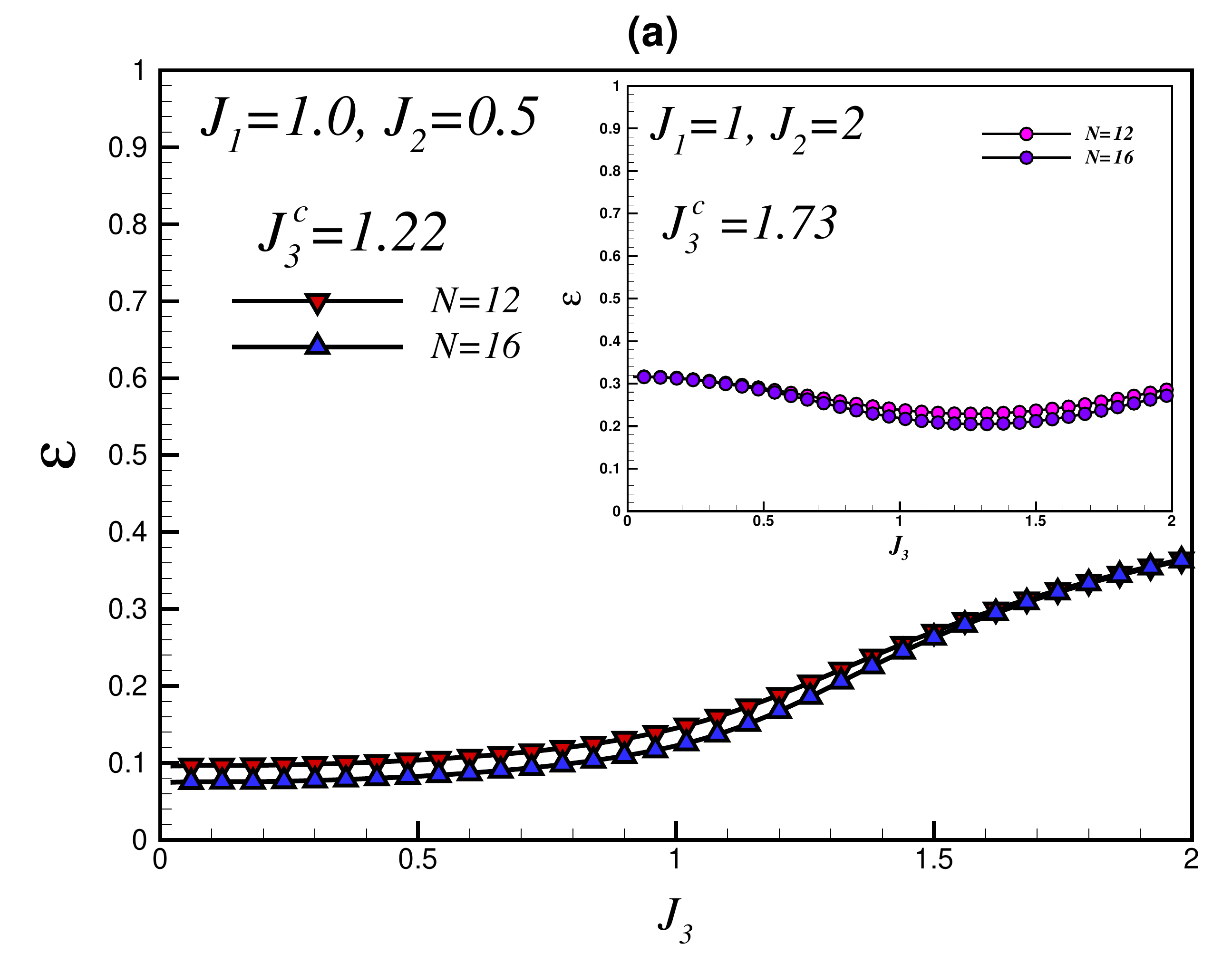}
\includegraphics[width=9cm]{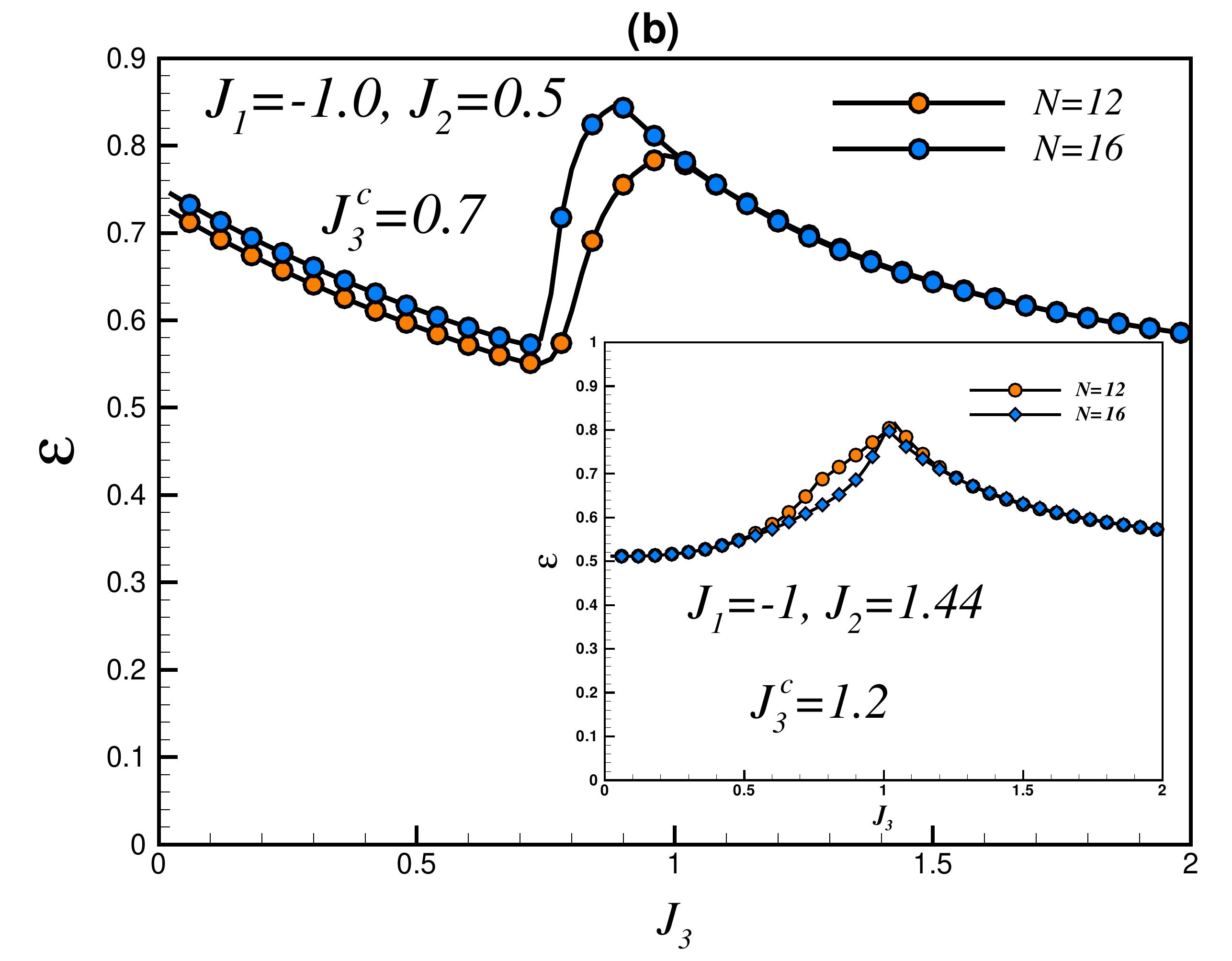}}
\caption{(Color online.) The numerical simulation results obtained from
the Lanczos algorithm for the geometric entanglement in finite chain sizes
$N=12, 16$. (a) For the exchanges $J_{1}=1, J_{2}=0.5$ and in the inset: $J_{1}=1, J_{2}=2$.
(b) For the exchanges $J_{1}=-1, J_{2}=0.5$ and in the inset: $J_{1}=1, J_{2}=1.44$.
} \label{fig9}
\end{figure*}

The numerical results of the geometric entanglement are plotted in Fig.~(\ref{fig9}) for the chain sizes $N=12, 16$. As is seen, the geometric entanglement is size-independent only in the cluster phase ($J_3>J_{3}^{c}$). The entanglement in the Neel phase is very small, but in the other regions of the  ground state phase diagram has a significant value. By increasing $J_3$, the entanglement remains almost constant up to the quantum critical point. At the critical value of the cluster interaction, a signature of the quantum phase transition is seen in Fig.~(\ref{fig9})(b). Which shows a dramatic change in the structure of the ground state of the system.

Topological order gives rise to a ground state degeneracy
that depends on the topology of the system and is
robust against any local perturbations \cite{Wen}. Because of this
property, topologically ordered systems appear to be good
candidates for robust quantum memory and fault-tolerant
quantum computation \cite{Kitaev}.
Not only can topological order explain exotic phases of
matter but it offers a whole new perspective to the problem
of elementary particles \cite{Alioscia}.

\begin{table}
\begin{center}
\label{table1}
\caption{The critical exponents $\theta$, $\tau$,  $\nu$. The scaling behavior of
the TPC function derivative in vicinity of the critical points.}
\begin{tabular}
{|c|c|c|c|c|} \hline
$exchange~~couplings$ & $Critical Points$  & $\theta$ & $\tau$ &  $\nu$\\
\hline
$J_{1}=1, J_{2}=0.8$ &$J_3^{c}=J_{3}^{0}=\sqrt{1.8}$ &$1.883$ &$0.15$   &$1.00$  \\
\hline
$J_{1}=1, J_{2}=2$   &$J_3^{c_{1}}=J_{3}^{\pi}=1$ &$1.907$    &$0.145$     &$0.985$  \\
\hline
$J_{1}=1, J_{2}=2$   &$J_3^{c_{2}}=J_{3}^{0}=\sqrt{3}$ &$1.896$ &$0.155$   &$1.00$  \\
\hline
$J_{1}=-1, J_{2}=0.19$  &$J_3^{c}=J_{3}^{\pi}=0.9$ &$2.524$ &$0.448$   &$0.989$  \\
\hline
\end{tabular}
\end{center}
\end{table}

\begin{figure*}
\centerline{\includegraphics[width=6.5cm,height=6.5cm,angle=0]{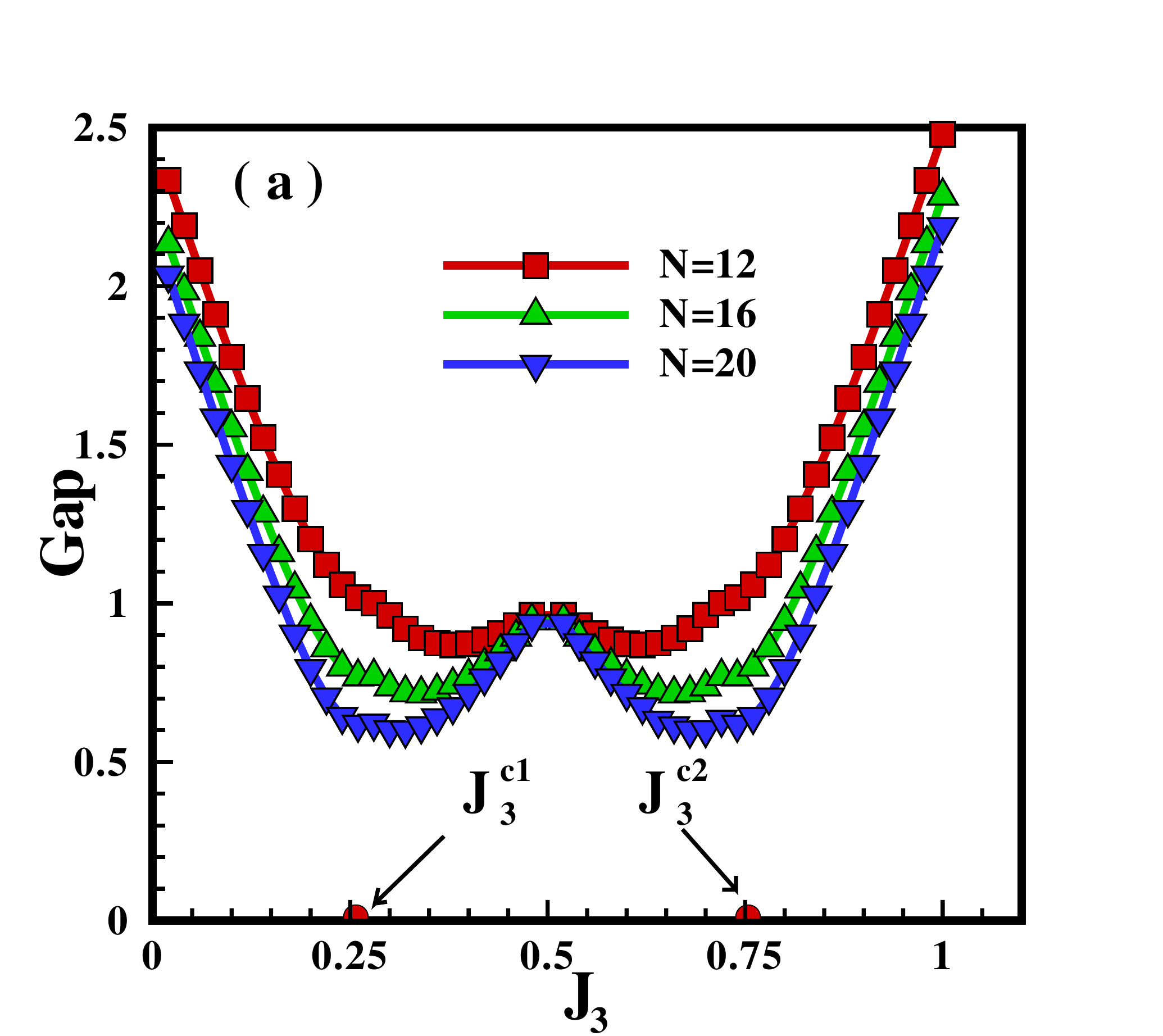}
\includegraphics[width=6.5cm,height=6.5cm,angle=0]{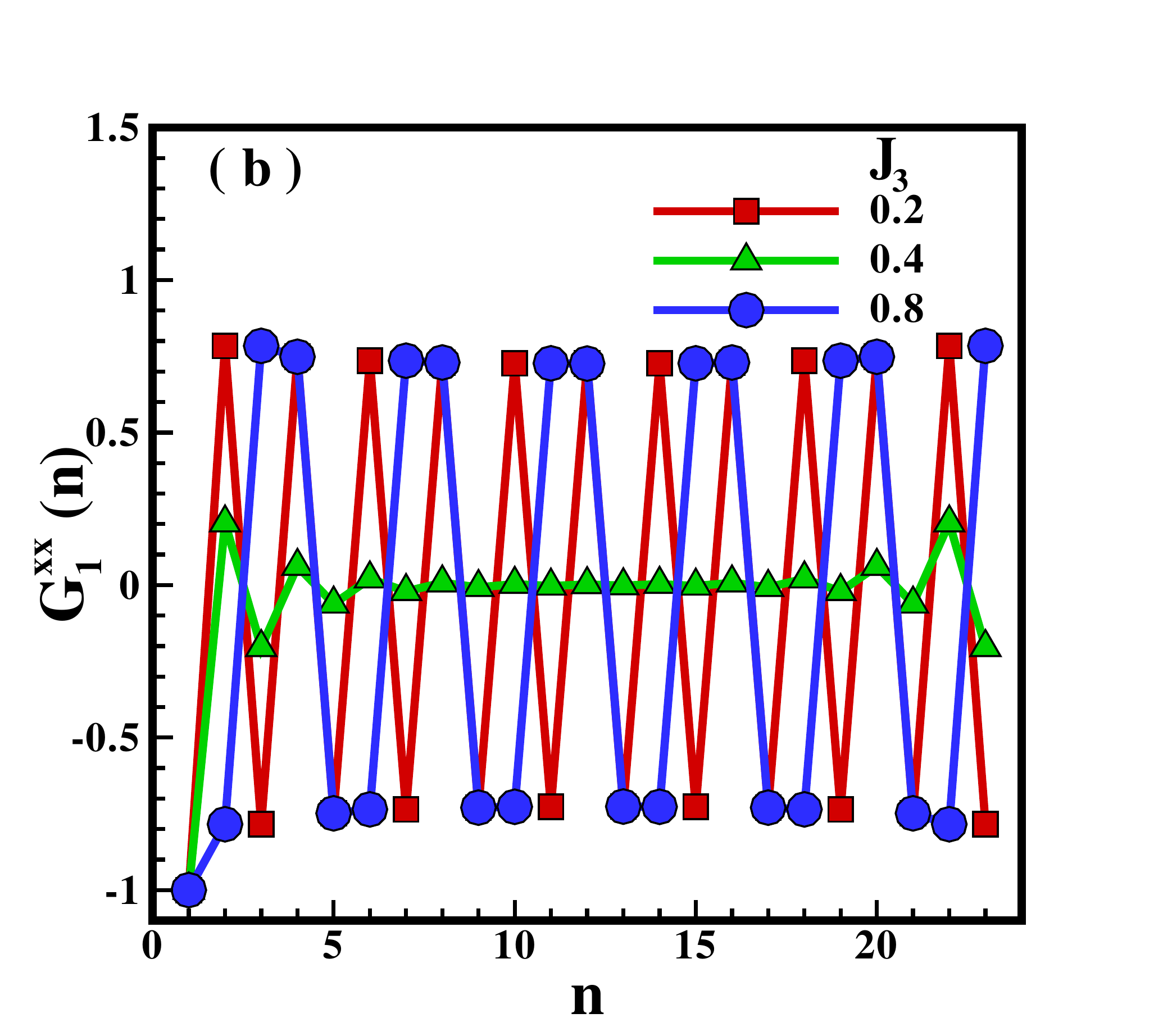}
\includegraphics[width=6.5cm,height=6.5cm,angle=0]{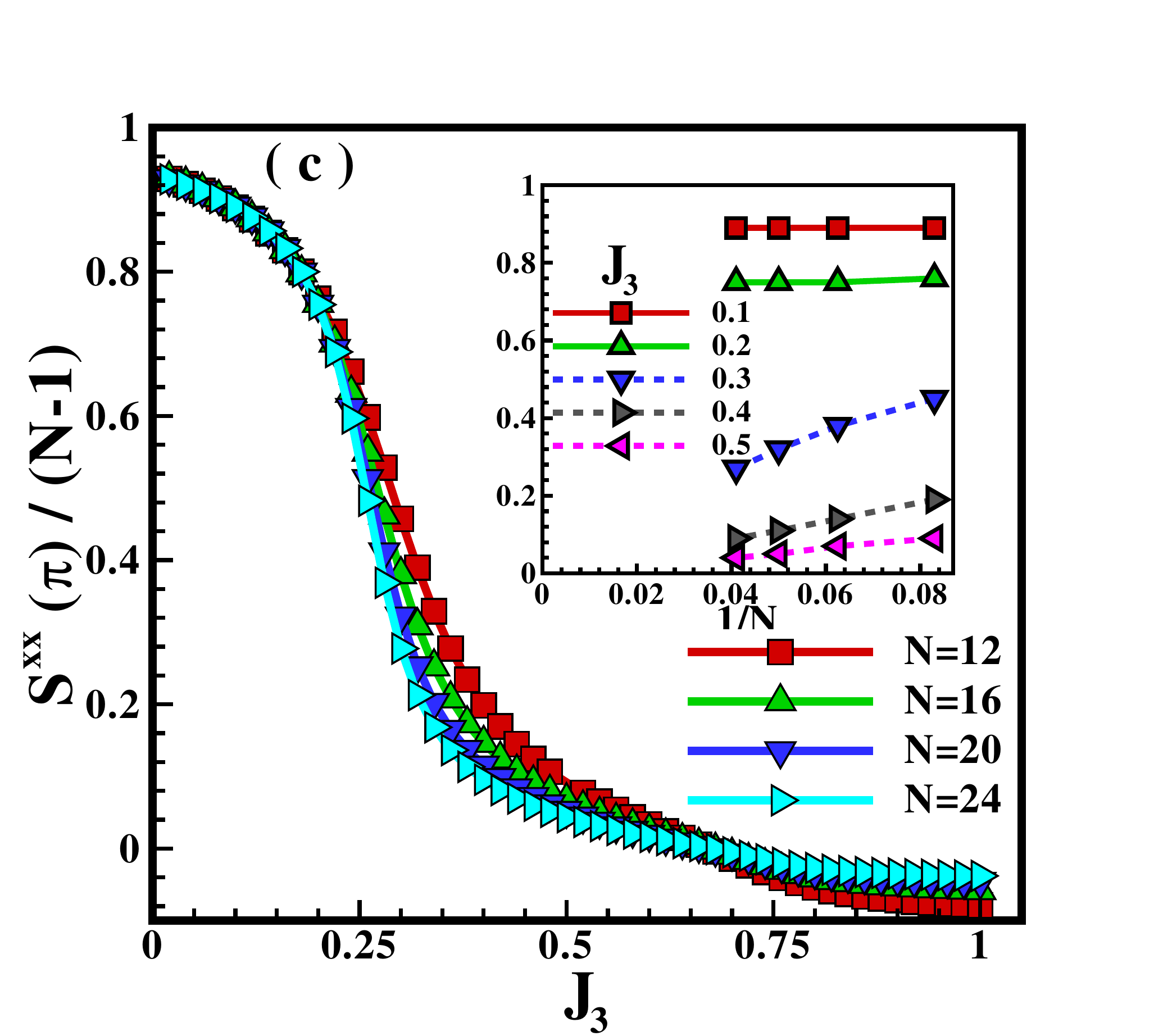}}
\caption{(Color online.) The numerical simulation results obtained from the Lanczos algorithm in the region $(J_1, J_2)=2.0, 0.5$. The energy gap is plotted versus the frustration $J_3$ in figure (a). It is seen that the energy gap is minimized at the two critical frustrations.  (b) The spin-spin correlation function, $W_1^{xx}(n)$, as a function of the $n$ for chain length $N=24$ and three values of the frustration corresponding to the different phases.  Figure (c) shows the spin structure factor at momentum $q=\pi$ versus $J_3$. In the inset, the spin structure factor $S^{xx}(\pi)$ is plotted as a function of the inverse chain length $1/N$ for different values of the frustration $J_3$. } \label{fig10}
\end{figure*}

\section{Frustrated Compass Model\label{USCF}}

In this section we consider the 1D frustrated quantum compass model with the Hamiltonian
\bea
\label{eq1}
H=\sum_{n=1}^{N'}[J_{1}\sigma^{x}_{2n-1}\sigma^{x}_{2n}+
J_{2}\sigma^{y}_{2n-1}\sigma^{y}_{2n}+ L_{1}\sigma^{x}_{2n}\sigma^{x}_{2n+1}
+J_{3} (\sigma^{x}_{2n-1}\sigma^{x}_{2n+1}+\sigma^{x}_{2n}\sigma^{x}_{2n+2})],
\label{Hamiltonian}
\eea
where $J_{3}$ denotes the usual NNN coupling. In following, using the numerical Lanczos method we study the effect of the frustration ($J_3$) in the ground state phase diagram of the QC model.
Due to the exponential growth of the Hilbert
space, numerical Lanczos simulation is limited to very small systems. Here we have
studied finite chains with $N=12, 16, 20, 24$ spins and periodic boundary conditions. The following quantities are computed: (i) the low-lying energies of the spectrum, (ii) spin-spin correlation functions and the spin structure factors, (iii) the Neel and stripe antiferromagnetic order parameters.


\subsection{Region I: $J_1>0$ and $J_2<1$}

The ground state of the QC model is known to be in the gapped Neel phase in the region ($J_1>0$, $J_2<1$)\cite{Mahdavifar}. To study the induced effects of the NNN interaction on the magnetic behavior of the ground state we did a very accurate simulation and the results are presented in Fig. \ref{fig10} for the values of exchanges $(J_1, J_2)=2.0, 0.5$. It is known that the energy gap is very informative and plays very important role in the quantum phase transition. Analyzing the numerical results of three lowest levels, we found the energy gap in finite chains should be considered as the difference between the energies of the ground and second excited states. From Fig. \ref{fig10}(a), it is clearly seen that the system is gapped at $J_3=0$ in good agreement with previous works\cite{Eriksson, Mahdavifar}. Also, the energy gap shows a universal behavior in respect to the frustration. As soon as the frustration is added the energy gap decreases and shows two minimums at certain values of the frustration. The location and value of minimums depend on the size system ($N$). In principle the minimum value of the gap decreases with increasing the chain size and using an extrapolation technique we found the gap will be closed in the thermodynamic limit $N$ at critical frustrations $J_3^{c_1}=0.27\pm 0.01, J_3^{c_2}=0.77\pm 0.01$. It should be noted that the 1D frustrated compass model in this region is gapped elsewhere. It means that by tuning the exchange interactions, the ground state of the system can be found in one of the three different gapped phases. Immediately, a question is arises about magnetic ordering of the system in these different gapped phases.

In numerical simulations, the best way to know the long-range magnetic order is the study of the spin-spin correlation
function defined by

\begin{eqnarray}
G_{j}^{\alpha\alpha}(n) = \langle\sigma_{j}^{\alpha}\sigma_{j+n}^{\alpha}\rangle~~~(\alpha=x, y, z),
\end{eqnarray}

and the spin structure factor at the momentum $q$ defined by

\begin{eqnarray}
S^{\alpha\alpha}(q) = \sum_{n=1}^{N-1}G_{j}^{\alpha\alpha}(n) exp(iqn).
\end{eqnarray}

\begin{figure}[b]
\begin{center}
\includegraphics[width=9cm]{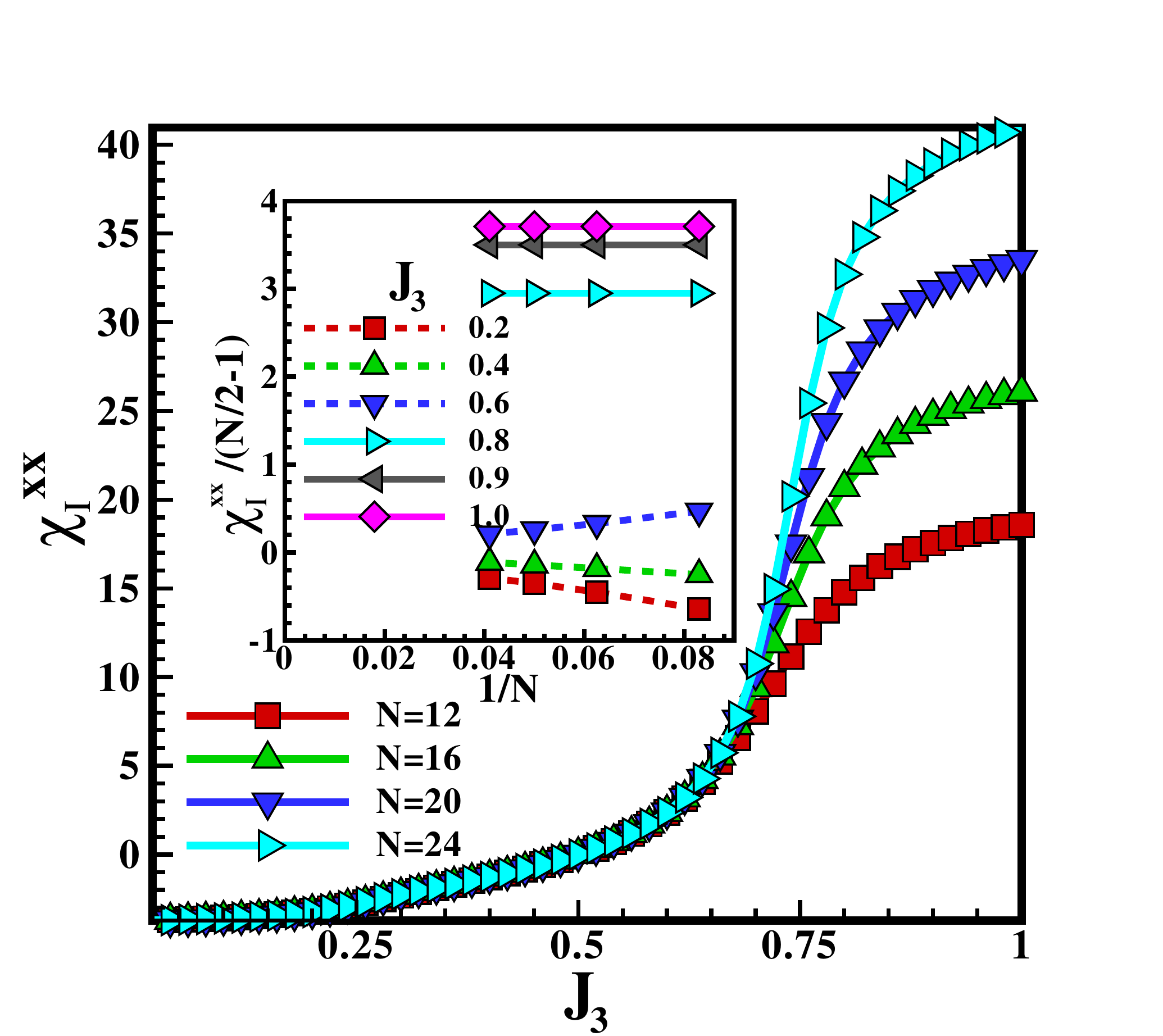}
\caption{(Color online) The correlation function of the
SAF-I order $\chi_{I}^{xx}$ as a function of $J_3$.
In the inset, the mean value of the correlation function
$\chi_{I}^{xx}/(N/2-1)$ as a function of $1/N$ for different
values of the frustration is shown.} \label{fig11}
\end{center}
\end{figure}

The spin structure factor gives us a deep insight into the characteristics of the ground state. In
Fig. \ref{fig10}(b) we have plotted $G_{1}^{xx}$ as a function
of $n$  for three different values of frustration $J_3=0.2<J_3^{c_1}$, $J_3^{c_1}<J_3=0.4<J_3^{c_2}$, $J_3=0.8>J_3^{c_2}$ and chain size $N=24$. It can be seen that the $x$ component of the
spins on odd sites are pointed in the same direction with the
$\sigma_{1}^{x}$ and others (on even sites) are pointed in
opposite direction  at $J_3=0.2$. This is an indication for
the N$\acute{e}$el ordering in the region $J_3<J_3^{c_1}$. In Fig. \ref{fig10}(c), we have plotted $S^{xx}(q=\pi)/(N-1)$ as a
function of $J_3$ which  is qualitatively the same as the staggered
magnetization, $M_{st}^{x}=1/N \sum_{j} (-1)^{j}\sigma_{j}^{x}$.

\begin{figure*}[t]
\centerline{\includegraphics[width=6.5cm,height=6.5cm,angle=0]{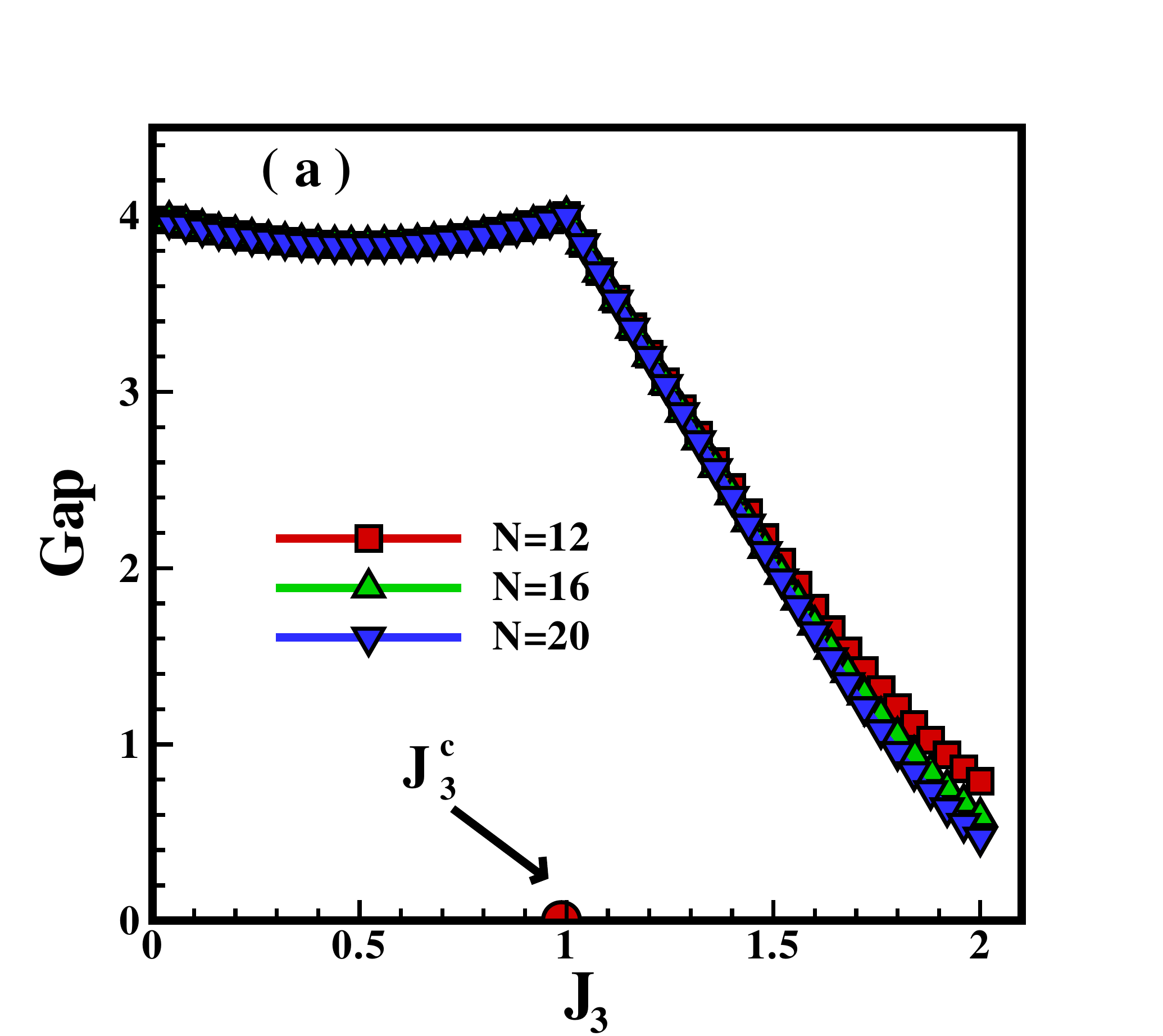}
\includegraphics[width=6.5cm,height=6.5cm,angle=0]{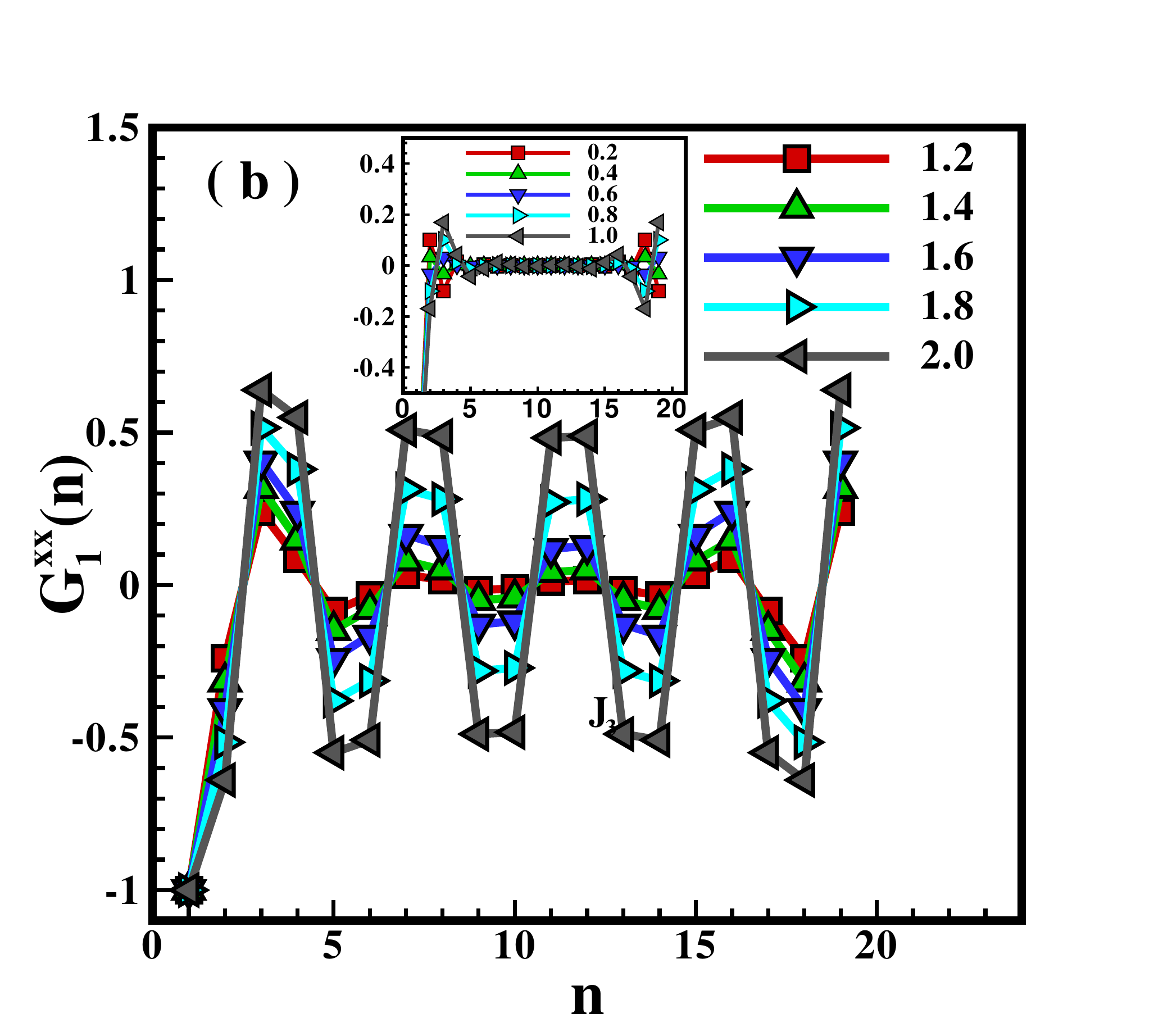}
\includegraphics[width=6.5cm,height=6.5cm,angle=0]{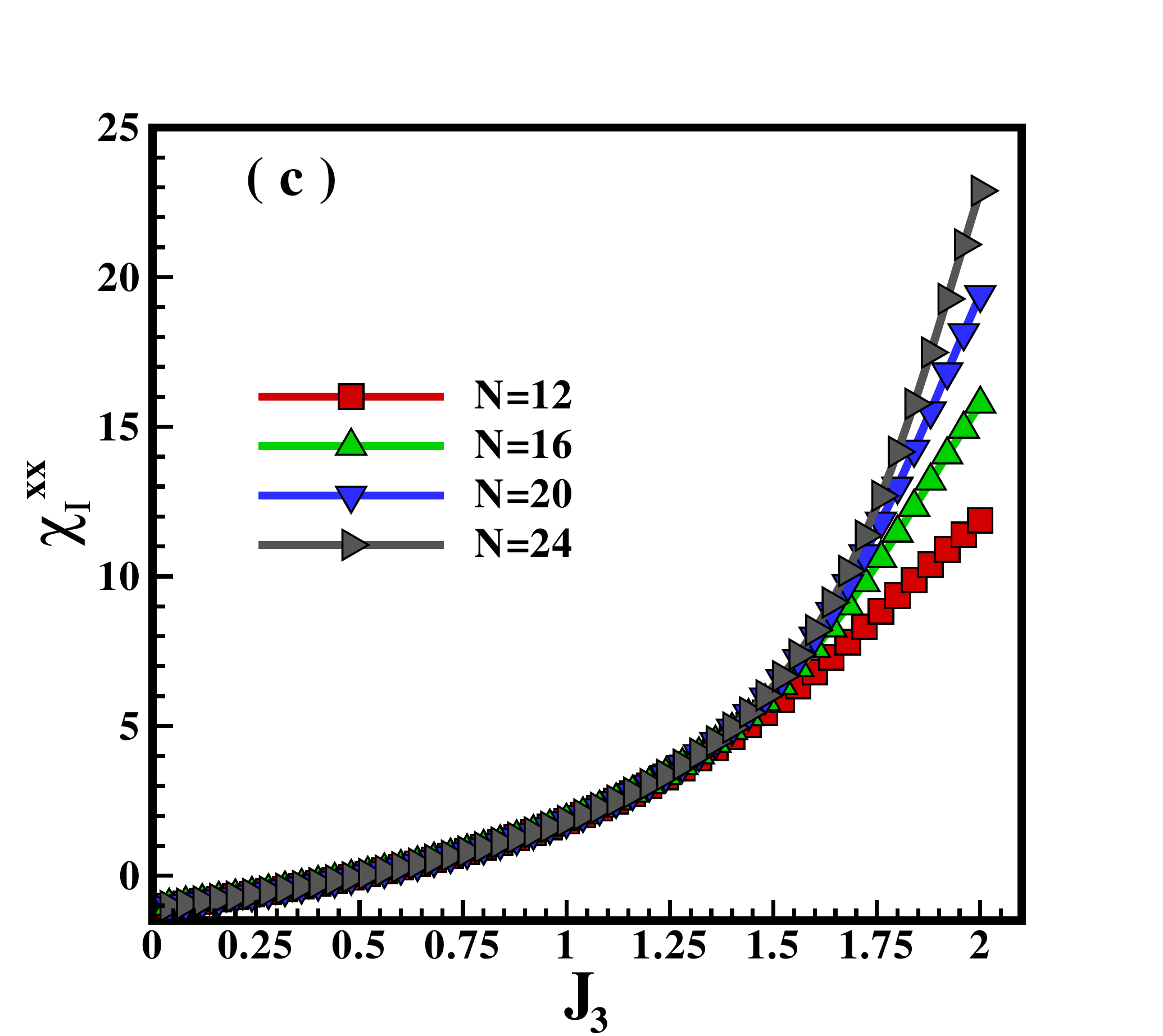}}
\caption{(Color online.) The numerical simulation results obtained from the Lanczos algorithm in the region $(J_1, J_2)=2.0, 3.0$. The energy gap is plotted versus the frustration $J_3$ in figure (a). It is seen that the energy gap shows two different behaviors in the regions $J_3<J_3^{c}$ and $J_3>J_3^{c}$.  (b) The spin-spin correlation function, $W_1^{xx}(n)$, as a function of the $n$ for chain length $N=20$ and different  values of frustration corresponding to $J_3>J_3^{c}$.  In the inset the same results are shown in the region $J_3<J_3^{c}$. Figure (c) shows The correlation function of SAF-I order, $\chi_{I}^{xx}$, as a function of $J_3$.  }
\label{fig12}
\end{figure*}

Numerical results, show that the un-frustrated system is in the saturated Neel phase. Adding the frustration the spin structure factor  $S^{xx}(q=\pi)/(N-1)$ decreases up to the first critical frustration $J_3^{c_1}$. By more increasing the frustration, $S^{xx}(q=\pi)/(N-1)$ drops down and a profound size effect is seen in the amplitude. In the inset of this figure, the effect of the size is checked. As is seen, the Neel ordering exist only in the region $J_3<J_3^{c_1}$. On the other hand, no long-range correlation in $x$-direction is not seen in the intermediate region $J_3^{c_1}<J_3=0.4<J_3^{c_2}$. We have to mention that the other components also do not show any correlation in the intermediate region. Finally, for the value of frustration  $J_3=0.8>J_3^{c_2}$, the behavior of the $x$ component of the correlation function suggests a different magnetic ordering. In this new phase, spins on even bonds are pointed in the same direction and those on odd bonds are pointed in the opposite directions. This is known as the stripe-antiferromagnetic-I phase\cite{Mahdavifar} and can be recognized from the stripe-antiferromagnetic-I (SAF-I) magnetization defined as

\begin{eqnarray}
M_{sp-I}^{x} = \frac{2}{N}\langle \sum_{j=1}^{N/2}(-1)^{j}(\sigma_{2j}^{x}+ \sigma_{2j+1}^{x}) \rangle.
\end{eqnarray}

Since in a finite system no symmetry breaking happens, the Lanczos results lead to zero value of $M_{sp-I}^{x}$. However we computed the correlation function of the SAF-I order parameter given by

\begin{eqnarray}
\chi_{I}^{xx} = \langle \sum_{n=1}^{N/2-1}(-1)^{n}(\sigma_{2j}^{x}+ \sigma_{2j+1}^{x}) (\sigma_{2j+2n}^{x}+ \sigma_{2j+1+2n}^{x}) \rangle.\nonumber\\
\end{eqnarray}

Numerical results of $\chi_{I}^{xx}$ are plotted in Fig. \ref{fig11}. The negative value of $\chi_{I}^{xx}$ in values of frustration $J_3<J_3^{c_1}$ is originated from the Neel phase in finite size systems. In the intermediate region, the small value of the $\chi_{I}^{xx}$ goes to zero in the thermodynamic limit $N\longrightarrow\infty$ (Inset of Fig. \ref{fig11}). But in the region $J_3>J_3^{c_2}$ a profound SAF-I order exists in the $x$ direction. By investigating the $N$ dependence of $\chi_{I}^{xx}/(N/2-1)$, we found that the mentioned SAF-I order is the long range (Inset of Fig. \ref{fig11}).


\subsection{Region II: $J_1>0$ and $J_2>1$}

It was found that the ground state of the 1D QC is in a gapped hidden order in this region of the exchange parameters. To draw a picture of the induced effects of the frustration we have implemented our Lanczos algorithm for different chain sizes $N=12, 16, 20, 24$ and different values of the exchanges. In Fig. \ref{fig12}, we have presented numerical  results for the values of the
exchanges parameter corresponding to $(J_{1},J_{2})= (2.0, 3.0)$.

\begin{figure*}[t]
\centerline{\includegraphics[width=9cm]{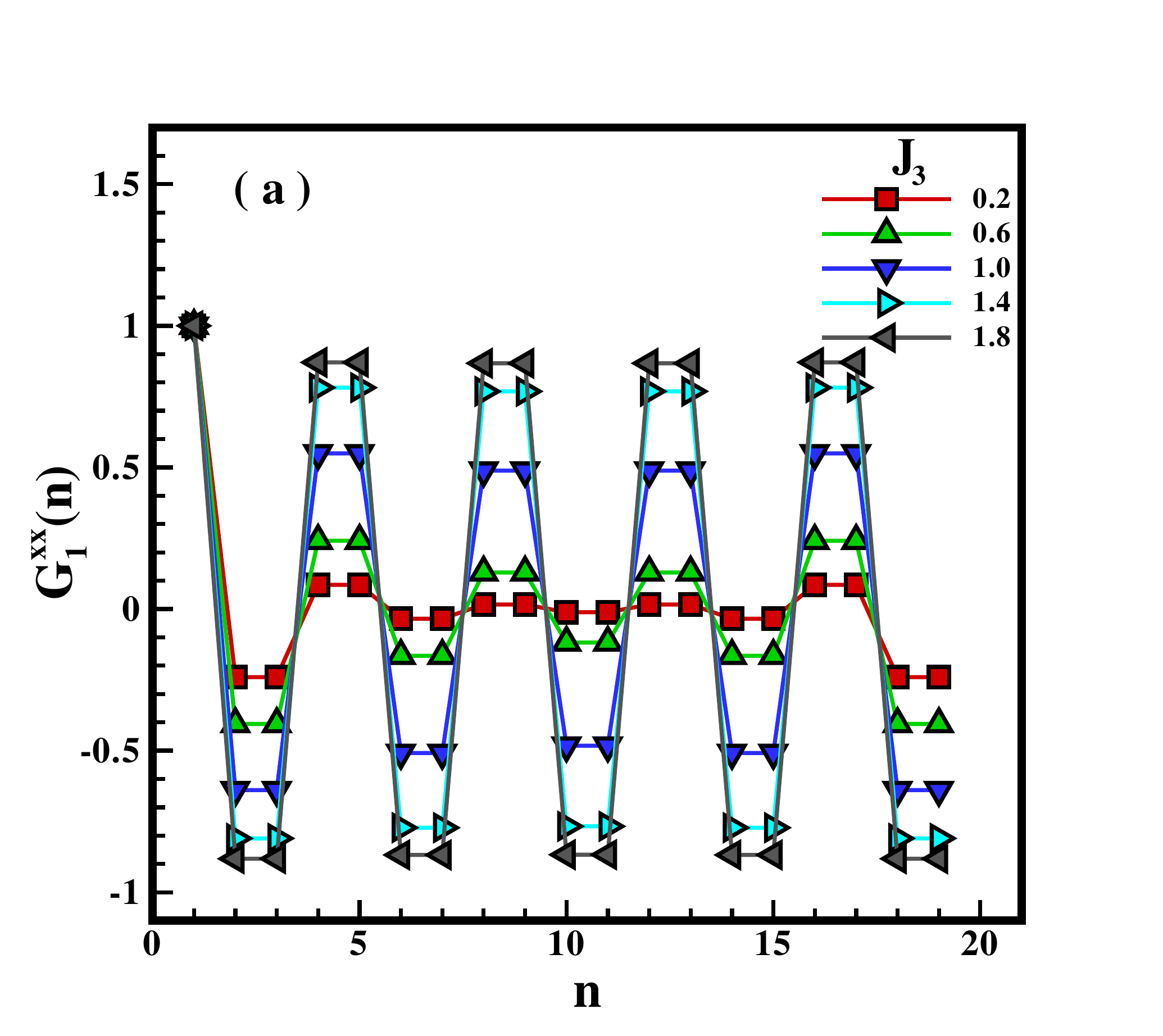}
\includegraphics[width=9cm]{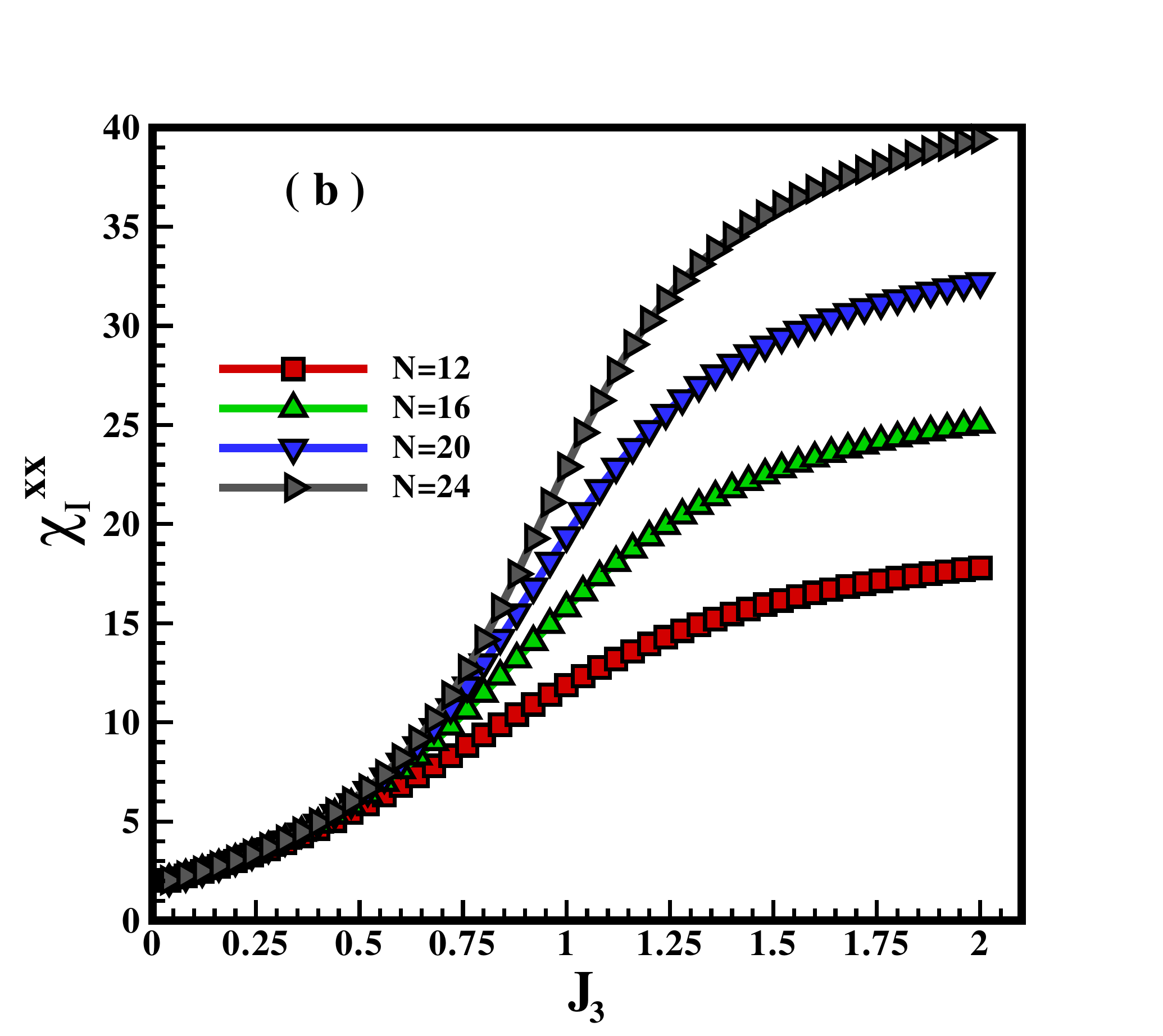}}
\caption{(Color online) The numerical simulation results obtained
from the Lanczos algorithm in the region $(J_1, J_2)=-2.0, 3.0$.
(a) The spin-spin correlation function, $W_1^{xx}(n)$,
as a function of the $n$ for chain length $N=20$ and different
values of frustration $J_3=0.2, 0.6, 1.0, 1.4. 1.8$.
Figure (b) shows the correlation function of SAF-I order
$\chi_{I}^{xx}$ as a function of $J_3$.}
\label{fig13}
\end{figure*}

In this region, we found the energy gap is characterized as a difference between the energies of the ground and first excited states. It is clearly seen from Fig. \ref{fig12}(a) that the spectrum of
the model is gapped in the absence of the frustration, $J_{3}=0$, in good agreement with previous works\cite{Eriksson, Mahdavifar} . Adding the frustration, the energy gap remains almost constant in the region $J_3<J_3^{c}$ but decreases rapidly as soon as the frustration becomes larger than the critical frustration $J_3^{c}$, which suggest  that the frustration has the ability to induce a new gapped phase in this region of the ground state phase diagram. To find the kind of the magnetic ordering in this new gapped phase we have calculated the spin-spin correlation functions. In
Fig. \ref{fig12}(b) we have plotted $G_{1}^{xx}$ as a function
of $n$  for values of frustration more than critical $J_3^c$.  It can be seen that the $x$ component of the
spins on even bonds are pointed in the same direction and those on odd bonds are pointed in
the opposite direction. Moreover, by increasing frustration, the amplitude of the correlation between spins on even bonds increases which shows that the frustration destroys quantum fluctuations. In addition, in the inset of Fig. \ref{fig12}(b), $G_{1}^{xx}$ is plotted for values of frustration $J_3<J_3^c$ where no long range order is seen. The behavior of the spin-spin correlation function shows that the suggested gapped hidden order in the 1D QC model\cite{Eriksson} will not surrender by adding the frustration and remains stable up to a critical frustration $J_3^{c}$. As we have mentioned the kind of induced ordering in the region $J_3>J_3^c$, is known as the SAF-I phase. Numerical results on, $\chi_{I}^{xx}$ are plotted in Fig. \ref{fig12}(c). The negative value of $\chi_{I}^{xx}$ at $J_3=0$ is originated from quantum fluctuations in hidden gapped order. By increasing the frustration $\chi_{I}^{xx}$ starts to increase but the overlapping data in the region $J_3<J_3^{c}$ is the indication of the short range correlations. In contrast, in the region of enough strong frustrations $J_3>J_3^{c}$, a
profound SAF-I order exists in the $x$ direction. By investigating the $N$ dependence of $\chi^{xx}_{I}$, we found that
the mentioned SAF-I order is a true long range order.


\subsection{Region III: $J_1<0$ and $J_2>1$}

It is known that the ground state of the 1D QC model is in a hidden gapped order phase in the region $J_1<0$ and $J_2>1$. Using our numerical simulation, we have implemented the Lanczos algorithm for chain sizes $N=12, 16, 20, 24$ and exchange parameters $(J_{1},J_{2})= (-2.0, 3.0)$. The numerical results of the correlation functions are plotted in Fig. \ref{fig13}. The spin-spin correlation function $G_{1}^{xx}$ is shown in Fig. \ref{fig13}(a). It can be seen that the NNN interaction induces the SAF-I correlations between spins as the same as with the region $J_1>0$ and $J_2>1$. Also, the amplitude of the correlation between spins on even bonds shows an increasing behavior with respect to $J_3$. In contrast to the previous hidden phase, a quasi long-range SAF-I order seems should be exist in very weak values of the frustration. In very recent works\cite{Jafari1, Motamedifar}, it was shown that the hidden orders in the ground state magnetic phase diagram of the 1D QC model show completely different behavior in a transverse magnetic field. Here, we also found the same qualitative behavior. In principle, hidden order in the region (II) will not surrender versus the NNN interaction, but in the region (III)  it will surrender as soon as the frustration is added.  To find a better picture of ground state magnetic phases of the system in this region, we have also plotted the $\chi_{I}^{xx}$ in Fig. \ref{fig13}(b). The positive value of $\chi_{I}^{xx}$ at very small values of the frustration can be originated from hidden order. By increasing the NNN interaction, $\chi_{I}^{xx}$ starts to increase and the overlapping data in the region $J_3\longrightarrow 0$ is the indication of the quasi long range correlations. In contrast, in the region of enough strong frustrations, a profound SAF-I order should be existed.


\subsection{Region IV: $J_1<0$ and $J_2<1$}

Classically, the effect of the negative exchange $J_1<0$ is
interesting. In the special case of $J_2=J_3=0$, the Hamiltonian
reduces to the alternating $XX$ Ising model. The ground state of
the alternating F-AF XX Ising model the long-range order canted spins in the
direction of the $x$ axis\cite{Mahdavifar}. The $x$-component of spins on odd bonds are pointed in the same direction and those on even bonds are pointed in the opposite direction. The ordering of this phase is called SAF-II phase. The order parameter of the SAF-II phase is defined as

\begin{eqnarray}
M_{sp-II}^{x} = \frac{2}{N}\langle \sum_{j=1}^{N/2}(-1)^{j}(\sigma_{2j-1}^{x}+ \sigma_{2j}^{x}) \rangle,
\end{eqnarray}
and the correlation function of the SAF-II order parameter given by

\begin{eqnarray}
\chi^{xx}_{II} = \langle \sum_{n=1}^{N/2}(-1)^{n}(\sigma_{2j-1}^{x}+ \sigma_{2j}^{x}) (\sigma_{2j-1+2n}^{x}+ \sigma_{2j+2n}^{x}) \rangle.
\end{eqnarray}

\begin{figure}[t]
\begin{center}
\includegraphics[width=9cm]{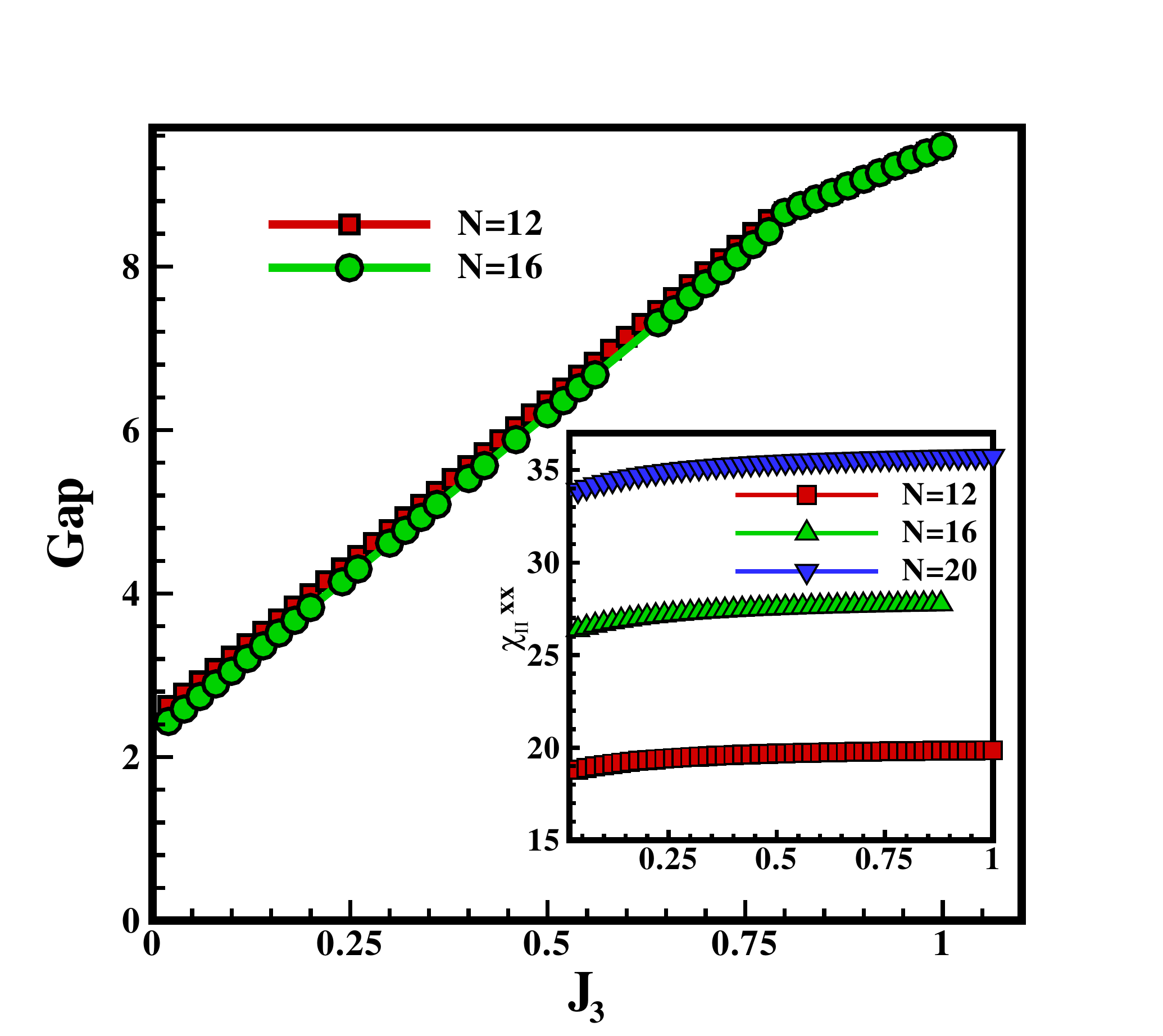}
\caption{(Color online) The numerical simulation results obtained from the Lanczos algorithm in the region $(J_1, J_2)=-2.0, 0.5$. The energy gap is plotted as a function of the frustration $J_3$. It is clearly seen that the energy gap decreases almost linearly by adding the NNN interaction and no phase transition does not observe. Inset, shows the correlation function of SAF-II order $\chi_{II}^{xx}$ as a function of $J_3$.}
\label{fig14}
\end{center}
\end{figure}

It has been shown that the induced quantum fluctuations by adding the exchange $J_2$ in this region cannot destroy the structure of the mentioned phase\cite{Mahdavifar}. Since the NNN interaction will not induce the frustration on the ground state of the system, we did not expect to find a quantum phase transition. The numerical results for exchange parameters $J_1=-2.0$ and $J_2=0.5$ and chain sizes $N=12, 16, 20, 24$ are plotted in Fig. \ref{fig14}. It is clearly seen that the system is gapped at $J_3=0$. Adding the NNN interaction, the desired gap grows linearly. In the inset, the correlation of the SAF-II, $\chi_{II}^{xx}$ is plotted as a function of the NNN interaction. It is completely clear that the long range order of the SAF-II phase is extended in total phase space.

\section{Conclusion}

In this work we have studied the difference between the induced quantum phases of the cluster interaction between next-nearest-neighbor (NNN) spins with the usual NNN interaction. In the first step, using the Jordan-Wigner transformation, an exact solution is obtained for the 1D quantum compass model with cluster interaction. We have obtained analytic expressions for critical couplings which drive quantum phase transitions. A rich quantum phase diagram including spin-flop, strip antiferromagnetic, antiparallel ordering of spin $x$ component on the leges and a magnetic phase with antiparallel ordering of spin $y$ component on rungs is obtained. In addition, the universality and scaling properties of the nearest-neighbor correlation functions derivatives
in different regions are studied to confirm the results obtained by the energy gap analysis.
In the second step, we have replaced the cluster interaction with the usual form of two point interaction. Using the numerical Lanczos method the Hamiltonian of the model is diagonalized for small chains up to $N=24$ spins. Analyzing numerical results, we have shown that the effect of the cluster interaction between NNN spins is completely different from the usual form. In fact, the Neel and two kinds of stripe antiferromagnetic phase (SAF-I and SAF-II) are found in the ground state phase diagram of the QC with added usual NNN interaction.

It would be interesting to study the model in the presence of a transverse field.
We have used the Jordan-Wigner transformation for searching the
phase diagram. We have been able to obtain the scaling behavior of
the NNC functions and NNNC functions versus the magnetic field.
The results is very interesting and show that there are several
nontrivial topological phase transitions in the extended cluster compass ladder
in a transverse field in different regions.
However, the quantum information properties (entanglement, concurrence,
quantum discord), quench dynamics and dynamics of entanglement of the model
has been studied in peresence/absent of the magnetic field.
As a consequence, we have found qualitative differences derived from the
nontrivial topological phase transitions.

\begin{acknowledgments}
The authors would like to thank S. A. Jafari, V. Karimipour,
H. Johannesson, G. I. Japaridze, R. Fazio for reading the manuscript
and valuable comments.
\end{acknowledgments}

\end{document}